\documentclass[preprint2]{aastex6}

\usepackage{multirow}
\usepackage{diagbox}

\newcommand{\CaK}{Ca\,II K}
\newcommand{\Hgamma}{$\rm{H}\gamma$}
\newcommand{\Hdelta}{$\rm{H}\delta$}
\newcommand{\HeI}{He\,I}
\newcommand{\HeII}{He\,II}
\newcommand{\MgII}{Mg\,II}
\newcommand{\NIII}{N\,III}
\newcommand{\SiIII}{Si\,III}
\newcommand{\NII}{N\,II}
\newcommand{\SiII}{Si\,II}
\newcommand{\SiIV}{Si\,IV}
\newcommand{\CII}{C\,II}
\newcommand{\OII}{O\,II}

\newcommand{\obsid}{\texttt{obsid}}

\shorttitle{LAMOST OB stars}
\shortauthors{Liu et al.}

\begin{document}


\title{A Catalogue of OB Stars from LAMOST Spectroscopic Survey}


\author{Zhicun Liu\altaffilmark{1,2}, Wenyuan Cui\altaffilmark{1}, Chao Liu\altaffilmark{2}, Yang Huang\altaffilmark{3},
Gang Zhao\altaffilmark{2}, Bo Zhang\altaffilmark{1}}
\affil{$^1$Department of Physics, Hebei Normal University, Shijiazhuang 050024, China;
wenyuancui@126.com}
\affil{$^2$Key Lab of Astronomy, National Astronomical Observatories, Chinese Academy of Sciences, Beijing 100101, China;
liuchao@nao.cas.cn}
\affil{$^3$South-Western Institute for Astronomy Research, Yunnan University, Kunming 650500, China}


\begin{abstract}
We present 22\,901 OB spectra of 16\,032 stars identified from LAMOST DR5 dataset. A larger sample of OB candidates are firstly selected from the distributions in the spectral line indices space. Then all 22\,901 OB spectra are identified by manual inspection. Based on a sub-sample validation, we find that the completeness of the OB spectra reaches about $89\pm22$\% for the stars with spectral type earlier than B7, while around $57\pm16$\% B8--B9 stars are identified. The smaller completeness for late B stars is lead to the difficulty to discriminate them from A0--A1 type stars. The sub-classes of the OB samples are determined using the software package MKCLASS. With a careful validation using 646 sub-samples, we find that MKCLASS can give fairly reliable sub-types and luminosity class for most of the OB stars. The uncertainty of the spectral sub-type is around 1 sub-type and the uncertainty of the luminosity class is around 1 level. However, about 40\% of the OB stars are failed to be assigned to any class by MKCLASS and a few spectra are significantly misclassified by MKCLASS. This is likely because that the template spectra of MKCLASS are selected from nearby stars in the solar neighborhood, while the OB stars in this work are mostly located in the outer disk and may have lower metallicity. The rotation of the OB stars may also be responsible for the mis-classifications. Moreover, we find that the spectral and luminosity classes of the OB stars located in the Galactic latitude larger than 20$^\circ$ are substantially different with those located in latitude smaller than 20$^\circ$, which may either due to the observational selection effect or hint a different origin of the high Galactic latitude OB stars.

\end{abstract}



\section{Introduction} \label{sec:intro}
Massive stars ($>8\,\mathrm{M}_\odot$), historically classified as OB stars, are the main source of chemical enrichment and re-ionisation of the universe, thus they play a major role in the evolution of their host galaxies. They can lead to various types of supernovae (SN), e.g. SN\,Ib, SN\,Ic, SN\,IIP \citep[e.g.,][]{poe08,pol97}. They may also result in a rich diversity of gamma-ray burst (GRB) phenomena, including long GRBs, soft GRBs et al. \citep[e.g.,][]{lan12}. Through powerful stellar winds and supernova explosions, massive stars can strongly influence the chemical and dynamical evolution of galaxies. They may also dominate the integrated UV radiation in high-redshift young galaxies.

Because of their enormous intrinsic brightness and short lives, massive OB stars ($>8\mathrm{M}_\odot$, referred as ``normal" OB stars hereafter) were usually used as valuable probes of present-day elemental abundances \citep[e.g.,][]{gie92,kil92,ven95,prz13}, especially for the blue supergiants referred as an ideal tracer of extragalactic metallicity \citep[see][and references therein]{kud08,kud12,kud14,kud16}. In addition, through studying the distances of about 30 OB stars in the solar vicinity, \citet{mor52} firstly published the Milky Way spiral structure sketch. Recently, \citet{xu2018} identified the distance from \emph{Gaia} DR2 data~\citep{gaia2018} for about 6\,000 bright OB stars \citep{reed2003} and map the nearby spiral arms with this samples. In addition, OB stars are also used as good tracers to explore the size of the Milky Way disk \citep{car10,car15,car17}, as they can be identified at large distance. Except the ``normal" OB stars, there are also some low-mass OB stars such as hot sub-dwarfs, white dwarfs, blue horizontal-branch (BHB) stars, post asymptotic giant branch (post-AGB) stars etc. 

Based on some stellar models, massive stars evolve through blue loops, that is, from main sequence to blue supergiant, to red supergiant, and then back to blue supergiant again \citep{sch92}. Hence, blue supergiant is an important phase and can tightly constrain the evolution models of massive stars \citep{mae14}. However, to date we still have poor knowledge of massive stars in the blue supergiant evolution stage, especially how strong their stellar wind affect the evolution. As a consequence, the final destination of blue supergiants is unclear. For instance, if they have sufficient mass loss, they can evolve into a Wolf-Rayet star and finally become a white dwarf, or explode as supernovae in Wolf-Rayet stage, or explode directly as supernovae in blue supergiant stage \citep{mas03,cro07,sma09}. Furthermore, the variable stellar winds have also been identified in some blue supergiants from their variable H$\alpha$ line profile, which made this issue more complicated \citep{aer10,kra15,hau16}. In addition, the mass loss during blue supergiant stage can play similar role to supernova remnant in the feedback from the massive stars to the ISM \citep{neg10}. 

Binarity may play also important role in the evolution of the massive stars \citep{lan08,min09,san13,san17}. In general, massive stars have high binary fraction \citep{mas09} and most of them belong to short period systems \citep{san11}. Because of the expanding of the evolved primary massive stars, mass exchanging will occur in close binaries between the primaries and their companions, which may drastically alter the evolutionary track of a massive star \citep{pod92,wel99,cla11,lan12}. 

Rotation is another important physical ingredient for massive stars, which may determine the fate of both single stars and binaries. \citet{dom03} pointed out that rotation deforms the star to an oblate shape, and extra mixing may be induced in the interior, which has been used to explain the observed surface abundance anomalies, such as nitrogen enrichment of several massive stars in main-sequence phase \citep[e.g.][]{heg00}. However, the existence of the highly nitrogen-enriched with slow rotation and relatively non-enriched with fast rotation found in the VLT-FLAMES survey samples \citep[see][and references therein]{min09} make the rotational mixing theory still in debate. 
These massive star samples with accurate abundance determinations and a wide range of rotational velocities were obtained from the VLT-FLAMES survey \citep{eva05,hun08}. 

Alternative processes, such as mass transfer in binaries, may also play some role in the nitrogen enhancements of massive main-sequence stars proposed by \citet{lan08}. In addition, rotation has been identified as an important factor to mass loss of massive stars for the evolution of both single star and binary \citep{mey00,mey05,heg00,lan12,eks12}.

To obtain new spectral classifications of at least all Galactic O stars brighter than $B = 13$, the Galactic O-Star Spectroscopic Survey (GOSSS) has been carried out to get high SNR blue-violet spectra with $R\sim2500$ for about 1000+ O stars within a few kpc of the Sun \citep{mai11}. \citet{sot11,sot14} presented spectral classifications for a total of 448 stars, which is almost the largest catalogue to date of the Galactic O stars with accurate spectral classification. In order to further study these O stars such as their binarity, physical parameters, and so on,  four other surveys (OWN, IACOB, NoMaDS, and CAF\'{E}-BEANS) have also been carried out to obtain high-resolution optical spectroscopy of a subsample of Galactic O stars in parallel to GOSSS \citep[][and references therein]{sot14}. In addition, the Bright Star Catalogue, also known as Yale Bright Star Catalogue, contains 9106 stars and 4 non-stellar objects, of which all stars with $V\leq6.5$ \citep{hol91}. This catalogue contains 50 O stars and 1711 B stars, which can be seen by naked eyes. However, there is no survey project on schedule mainly focuses on Galactic B stars. \citet{rom18} developed a near-infrared semi-empirical spectral classification method and successfully identified four new O stars, which was mistakenly classified as later B stars before, from the DR14 APOGEE spectra of 92 known OB stars.

While ionized helium (\HeII) lines appear in O stars , B stars are mainly identified from the optical neutral helium (\HeI) lines, which will essentially disappear in the spectra of A stars and O stars~\citep{gra09}. Spectral analyses with large and homogeneous samples are very valuable for OB stars, because some new observational constraints can be outlined on their origin and evolutionary status. Furthermore, the distant and faint OB stars in the Milky Way are still rare. In this paper, we identify more Galactic OB stars from the data release 5 (DR5) of the Large Sky Area Multi-Object Fiber Spectroscopic Telescope (LAMOST) survey, which contains around 8 million stellar spectra. Because of this survey mainly focus on the anti-center direction of the Milky Way, more distant and faint Galactic OB stars in the Galactic outer disk are identified in this work. Because their lower metallicities than the ones in the solar neighborhood, these new OB stars may provide different evolution tracks and thus give new observational constraints.

The paper is organized as the following. In section 2, we briefly introduce the basic data from LAMOST DR5, and the details of the method to search for OB stars is descripted in section 3. In section 4, the sub-classification of the identified OB stars is conducted. In section 5, we summarize the distribution of different sub-types of OB stars in line indices space and in spatial locations. Finally, a short conclusion is drawn in section 5.

\section{Data}
\subsection{The LAMOST Data}\label{sect:data}

The LAMOST, also called Guo Shou Jing telescope, is a 4-meter reflective Schmidt telescope with 4000 fibers on a 20-square degree focal plane and a wavelength coverage of 
370-900\,nm \citep{cui12,zha12,luo12}. The unique design of LAMOST enables it to take 4000 spectra in a single exposure to a dynamical range of magnitude from $r=9$ to $18$\,mag at the resolution R\,=\,1800~\citep{den12}. The LAMOST survey has finally obtained more than 9 million low resolution stellar spectra, 8 million of which are stellar spectra, after its 5-year survey (DR5). Previous works have shown that LAMOST observations are biased more to the giant than the dwarf stars, since the limiting magnitude is relatively bright \citep{liu14a,wan15}. With such a large amount of spectra, we are able to identify many interesting but relatively rare stellar objects, such as carbon stars~\citep{ji2016,li2018}, Mira variables~\citep{yao2017}, or early type emission line stars~\citep{hou2016} etc.

\subsection{Line Indices}\label{sect:lineind}

According to \citet{liu15,liu17}, the selection function of LAMOST survey does not strongly bias to any spectral type of star. This implies that LAMOST catalog should be composed of almost all types of stars. \citet{liu15} suggested to classify all types of normal stars from spectral line indices. Compared to the traditional approach to classify the spectral types, using line indices allows a semi-automatic classification. With multiple spectral line indices, various spectral types may transitionally change from one to another, which naturally reflect the variation of the astrophysical parameters, e.g. effective temperature, surface gravity, and luminosity, from type to type. 

The line index in terms of equivalent width (EW) is defined by the following equation \citep{wor94,liu15}:
\begin{equation}
EW=\int(1-\frac{F_\lambda}{F_C})d\lambda,
\end{equation}
where $F_\lambda$ and $F_C$ are the fluxes of spectral line and pseudo-continuum, respectively. $F_C$ is estimated via linear interpolation of the fluxes located in the ``shoulder'' region on either side of the line bandpass. The unit of line index under this definition is in \AA. For the spectra with signal-to-noise ratio larger than 15, the typical uncertainty of the equivalent widths of the atomic lines is smaller than 0.1\,\AA\, \citep{liu15}.

In light of \citet{liu15} we identify LAMOST observed OB type stars from the selection in line indices space. In general, as the hottest normal stars, OB type stars show moderately weak Balmer lines with strong neutral or ionized He lines. The neutral metal lines, however, are very weak or even disappeared. We apply these features to discriminate OB type stars from other cooler ones.

We re-calculate the equivalent width of spectral lines listed in Table\,2 of \citet{liu15} for the LAMOST DR5 spectra. Moreover, we additionally calculate the equivalent width of \CaK\ line following the definition that the line bandpass is at ($3924.7$, $3942.7$)\,\AA\ and the two continua bands are at ($3903$, $3923$) and ($4000$, $4020$)\,\AA\ following \citet{beers1999} and the equivalent width of \HeI\ (4471\,\AA) providing that the line bandpass is ($4462$, $4475$)\,\AA\ and the left and right continua are ($4450$,$4463$) and ($4485$, $4495$\,\AA), respectively. To obtain a stable measurement of Fe, we average over all nine Fe lines defined in Lick indices \citet{wor94}. That is, we denote $EW_{Fe}$ as the mean value of lines $4383$, $4531$, $4668$, $5015$, $5270$, $5335$, $5406$, $5709$, and $5782$\,\AA. 
 
\section{Identification of OB stars}

We firstly select the stellar spectra with signal-to-noise ratio (S/N) larger than $15$ in $g$-band from LAMOST DR5, and obtain 4\,940\,840 stellar spectra. It should be noted that some stars have been observed multiple times and contribute several spectra.

\begin{figure}
	\plotone{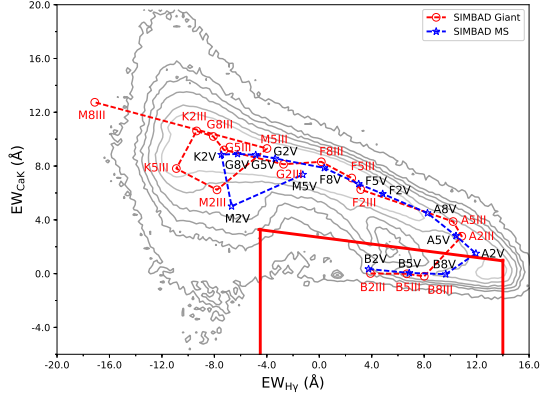}
	\caption{Distribution of the LAMOST DR5 stars with $\mathrm{S/N}>15$ at $g$-band 
	(the contours) in the \CaK\ (3933\,\AA) vs. \Hgamma\ (4340\,\AA) plane. The blue dashed line with the asterisks and red dashed line with the unfilled circles represent the loci of the main-sequence and giant stars provided by \citet{liu15}, respectively. The red solid lines are the cuts for OB candidate, following Equations (\ref{eq:cri1_1}) and (\ref{eq:cri1_2}).\label{fig:cak-hgamma}}
\end{figure}

Second, we map the above stars in \CaK\ vs. \Hgamma\ plane (see Figure~\ref{fig:cak-hgamma}). We overlap the loci of the main-sequence and giant stars provided by \citet{liu15} and find that OB stars are mostly located at the area with smaller values of \Hgamma\ and \CaK. Therefore, we empirically select the area surrounded by the red solid lines such that most of the stars following the loci of the OB types are included. Specifically, we select the stars satisfying the following criteria
\begin{equation}\label{eq:cri1_1}
\mathrm{EW_{CaK}}<2.5-\mathrm{EW_{H\gamma}}/8,
\end{equation}
and
\begin{equation}\label{eq:cri1_2}
-4.5\leq\mathrm{EW_{H\gamma}}\leq14.
\end{equation}
After applying this cut, 151\,902 OB candidates are selected.

\begin{figure}
	\plotone{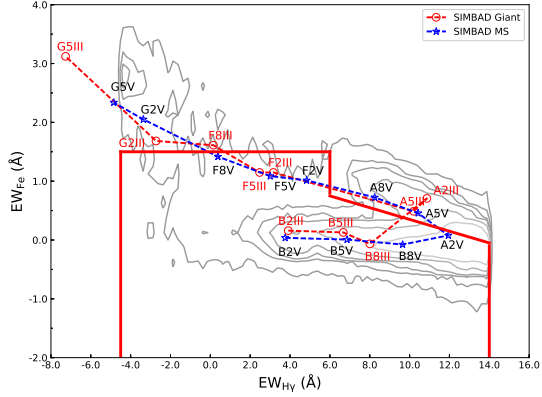}
	\caption{Distribution of 151\,902 OB star candidates (the contours) in the Fe vs. H$\gamma$ plane. The symbols are same as in Figure~\ref{fig:cak-hgamma}. The red solid lines, which are based on Equations (\ref{eq:cri1_3}), are the line cuts used to further select OB star candidates. \label{fig:Fe-Hgamma}}
\end{figure}

Third, we further map the candidate samples in the Fe vs. \Hgamma\ plane and remove the stars with strong Fe absorption features (see Figure~\ref{fig:Fe-Hgamma}). Although the figure indicates the location of the late type stars, not many late type stars are in the 151\,902 samples after the previous cuts showing in Figure~\ref{fig:cak-hgamma}. The following empirical criteria is adopted 
\begin{eqnarray}\label{eq:cri1_3}
	\mathrm{EW_{Fe}}<1.5&\text{ if } \mathrm{EW_{H\gamma}}<6,\nonumber\\
	\mathrm{EW_{Fe}}<-0.1 (\mathrm{EW_{H\gamma}}-6)+0.75&\text{ if }\mathrm{EW_{H\gamma}}\ge6.
\end{eqnarray}
This cut can exclude most of the late type stars, according to the stellar loci overlapped in the figure. We allow the cut in Fe slightly larger at \Hgamma$<6.0$ to include most of the giant and supergiant stars since they may located in between the locus of the B type main-sequence and that of late type stars. We finally obtain 122\,504 OB candidates, in which 19\,681 stars are with \Hgamma\,$<6$ and 102\,823 stars with \Hgamma\,$\geq6$. The OB candidates with \Hgamma\,$\geq6$ are substantially contaminated by A, F and G stars, which have very weak or no \HeI\ absorption features.

It is noted that the equivalent widths of \CaK, \Hgamma, Fe and \HeI\ have negative values, which are due to the variation of the pseudo-continua affected by other absorption lines. This systematics is very difficult to be removed since the nearby strong absorption lines make it impossible to find the real continua. Nevertheless, the systematic offsets would not affect the identification of the weak \CaK, Fe, and \Hgamma\ features and strong \HeI\ (4471\AA) lines, because not the absolute values of EWs, but the relative values play key role in the identification of OB stars. 

In addition, using the above selection criteria we may miss the stars without \CaK, \Hgamma, or Fe measurement. This is mostly due to the spectra with bad fluxes in the relevant wavelength regions. To identify as many as OB stars, we add additional 2\,055 stars with incomplete measurements of any of the above lines to the OB star candidates.

We further inspect the spectra for totally 124\,559 (122\,504+2\,055) OB star candidates by eyes for confirmation. To judge whether the candidate spectra are real OB stars, we mainly used the following line features, such as the \Hgamma, \Hdelta, \HeI\ lines at 4026, 4387, 4471\,\AA, \HeII\ lines at 4200, 4541, 4686\,\AA, \MgII\ at 4481\,\AA\, \SiIII\ triplet at around 4552\,\AA, and \NIII\ triplet at 4634, 4640, 4642\,\AA, etc. The spectra without He\,I and He\,II line features are firstly removed. Then, the intensity of the Balmer lines are considered as the indicators of spectral type, because they are prominent and reach the maximum at A2. The ratios of the strengths of He\,I to He\,II lines are also used to distinguish O and B stars, because He\,II lines usually absent in the spectra of B stars and the intensities of He\,I tends to increase with decreasing temperature while He\,II decreases. The ratio He\,I\,$4471/\mathrm{Mg\,II}\,4481$\AA\ is used to diagnose the contaminated stars, when it is less than 0 the star is removed.
This procedure has been done at least twice by the same people and has been independently confirmed by other people to avoid severe bias of the judgment. 

Finally, we identify 22\,901 spectra of 16\,032 OB stars from the 124\,559 spectra of the candidate sample above via the visual inspection. The final OB sample includes 135 spectra of 91 O stars, 21\,658 spectra of 15\,087 B stars, 948 spectra of 727 hot subdwarfs (including sdO and sdB), and 160 spectra of 127 white dwarfs (WDs). The summary of the numbers are listed in Table~\ref{tab:numbers} and the catalog is listed in Table~\ref{tab:OBstars}. Because in this work we only concern about the ``normal" OB stars, the hot subdwarfs and WDs selected in this work aren't for purpose and thus are not guaranteed to be the complete samples.

\begin{figure}
	\epsscale{0.85}
	\plotone{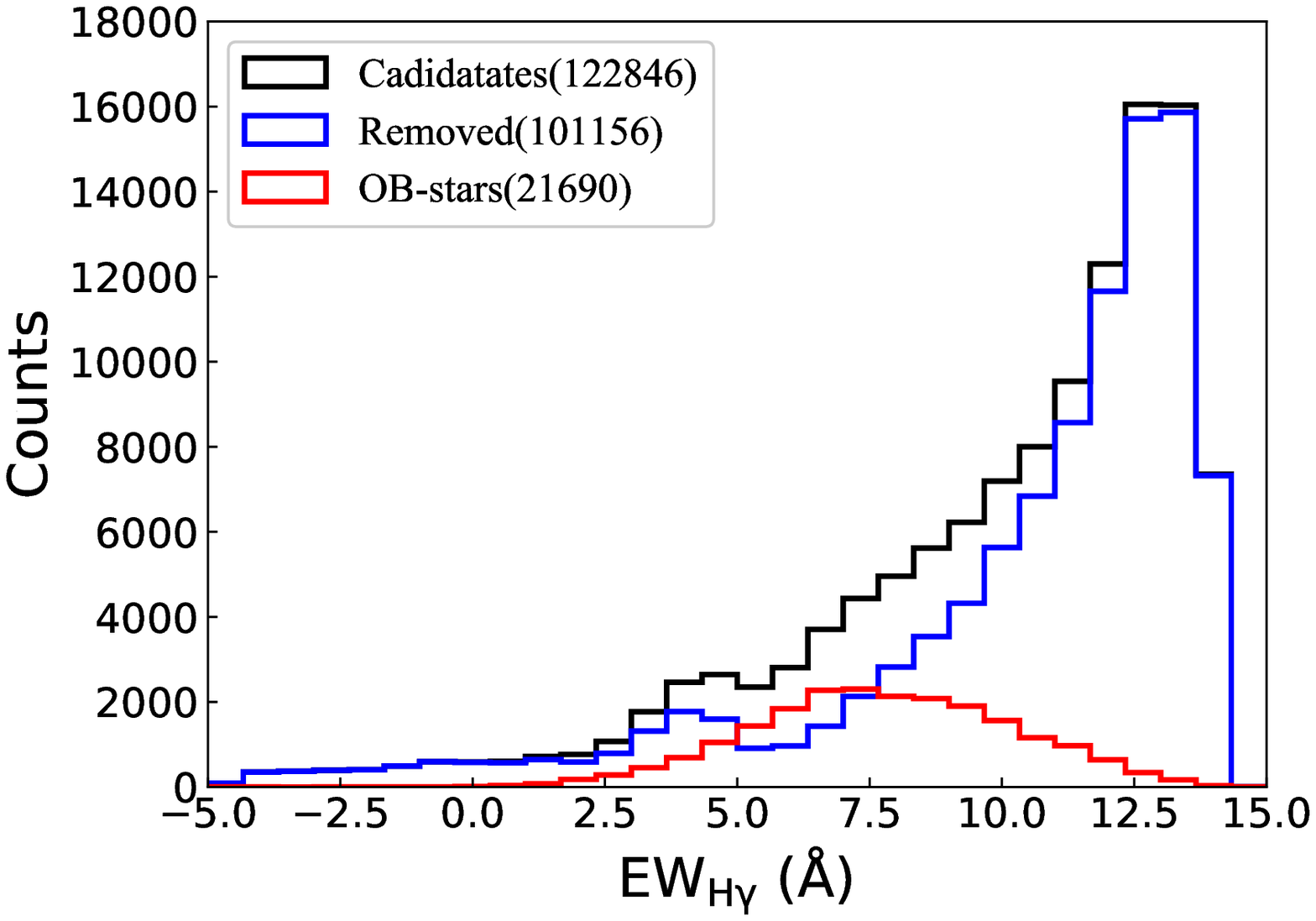}
	\plotone{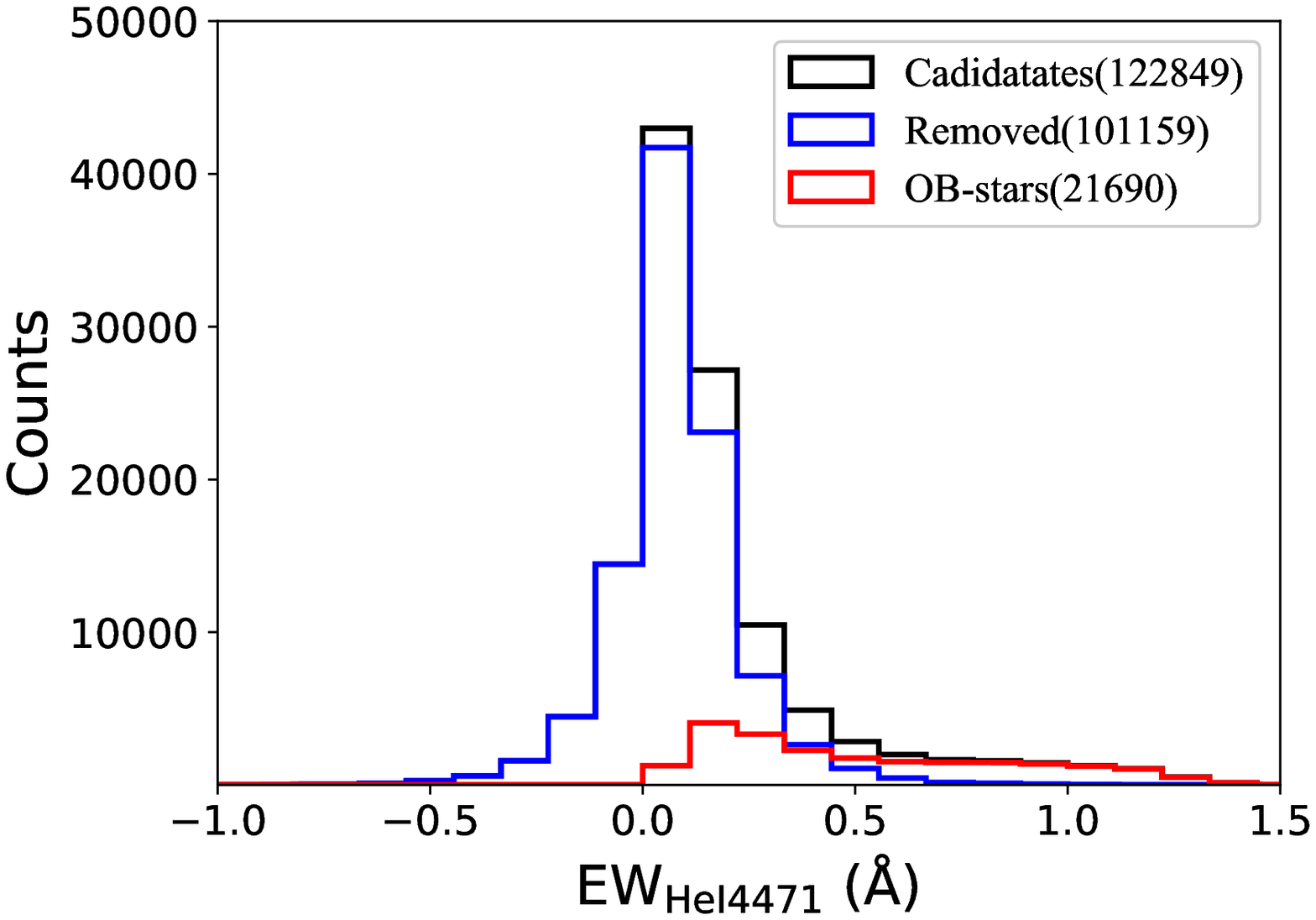}
	\caption{Top panel: Histogram of the EW of H$\gamma$\,(4340\,\AA)
	for 122,846 OB star candidates (black) and 21,690 identified normal OB stars (red) and 101,156 non-OB (blue) stars. 
	Bottom panel: Histogram of the EW of He\,I\,(4471\,\AA) for 122,849 OB star candidates (black) and 21,690 identified normal OB stars (red) and 101,159 non-OB (blue) stars.
	\label{fig:hgamma-HeI-hist}}
\end{figure}

\begin{table}
\caption{Numbers of OB stars.}\label{tab:numbers}
\begin{tabular}{l|c|c}
\hline\hline
Class & No. of spectra & No. of stars\\
\hline
	OB candidates$^a$ & 124\,559 & - \\
	normal O stars & 135 & 91 \\
	normal B stars & 21\,658 & 15\,087\\
	normal OB stars$^b$ & 21\,793 & 15\,178\\
	hot subdwarfs & 948 & 727 \\
	white dwarfs & 160 & 127 \\
	Total OB stars$^c$ & 22\,901 & 16\,032\\
\hline\hline
\end{tabular}
\\	
a: The candidates contains 122\,504 spectra selected from spectral line indices and 2\,055 without line measurements.\\
b: This is the sum of normal O and B stars (the above two rows).\\
c: This is the total number of normal OB, hot subdwarfs and white dwarfs (the sum of the above three rows). 
\end{table}

The top panel of Figure~\ref{fig:hgamma-HeI-hist} shows the histograms of \Hgamma\ (top panel) for the 21\,690 ``normal" OB stars, the excluded 101\,156 non-OB stars, and the total 122\,846 candidates, respectively. It is noted that the 948 hot subdwarfs and 160 white dwarfs are not included in this figure. Besides, 103 of 21\,793 (135 O stars+21\,658 B stars) ``normal" OB stars and 605 of the total 124\,559 candidates with no measurements of the equivalent widths are not included in this figure. 
The bottom panel shows the distribution of \HeI\ (4471\,\AA, bottom panel) for the same three groups, but the stars without measured equivalent width of \HeI\ (4471\,\AA) are not included. 
It is seen that many of (but not exclusively) the visually-selected OB stars are found in the region of strong \HeI\ ($>0.1$) and moderately strong \Hgamma\ (peak at about $7$). However, 11\,247 non-OB stars are also located in the moderately strong \Hgamma\ ($5.0-10.0$) and strong \HeI\ ($>0.1$) regime. Their fraction is about 43.7\% of 25\,703 OB candidates in this regime. They are double checked and found that they are contaminators due to the noise in the spectra. Some of the moderately strong \Hgamma\ stars also show strong \HeI, most of them are not OB stars but their \HeI\ measures are affected by the local spikes in the spectra. 

\subsection{Completeness}
\begin{figure*}
	\plotone{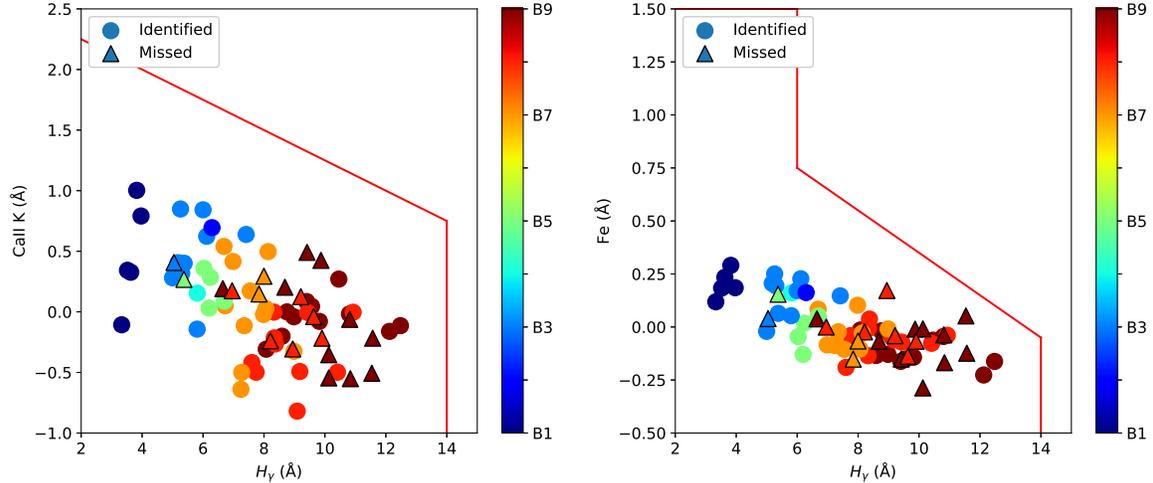}
	\caption{The left and right panels display the \CaK\ vs. \Hgamma\ and Fe vs. \Hgamma\ distribution of the OB stars found in an arbitrarily selected 5000 spectra with S/N$>15$ and within Galactic latitude of 20$^\circ$. The filled circles are OB stars identified with the previous approach, while the filled triangles are those failed to be identified. The colors code the spectral sub-types of these spectra according to the criteria of Table~\ref{tab:subclass}. The red lines in both panels stand for the same selection criteria used in Figures~\ref{fig:cak-hgamma} and \ref{fig:Fe-Hgamma}, respectively.}\label{fig:completetest1}
\end{figure*}
\begin{figure}
	\plotone{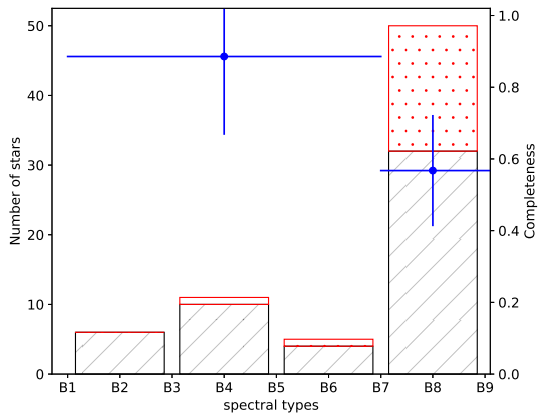}
	\caption{The black and red histograms show the distribution of spectral sub-types for the identified and missed OB stars, respectively, in the arbitrarily selected 5000 spectra. The two blue dots align with the right hand-side y-axis, which indicate the completeness of the selection criteria. The left blue dots indicates the completeness for the stars earlier than B7, while the right one indicates the completeness for the stars later than B7.}\label{fig:completetest2}
\end{figure}
We arbitrarily select 5000 spectra with S/N$>15$ and Galactic latitude within (-20$^\circ$ and +20$^\circ$) from LAMOST DR5 and manually inspect the spectra. We identify 72 OB type spectra from the random sample, 52 of which are in our list previously identified and 20 have been missed. This means that the completeness overall spectral types is around 72\%. 

We assign the spectral sub-type by eye inspection for these spectra according to the criteria listed in Table~\ref{tab:subclass}. We then map both the previously identified and missed spectra in the \CaK\ vs. \Hgamma\ and Fe vs. \Hgamma\ planes in Figure~\ref{fig:completetest1}. It shows that all the identified and missed OB stars comply with the selection criteria described by Equations~(\ref{eq:cri1_1})--(\ref{eq:cri1_3}). This means that the first cuts in the line indices space for OB stars are quite robust and essentially do not miss any OB stars. 

Figure~\ref{fig:completetest2} shows the histograms of the spectral sub-types for the identified and missed OB stars. Based on this small sub-sample, the completeness of the identified OB stars earlier than B7 is better than $89\pm22$\%, while the completeness of the OB stars later than B7 is larger than $57\pm16$\%. Note that the missed OB stars are all later than B6. This is probably because that the late B type stars are quite difficult to be discriminated from the early A type stars because weak metal line features are sometimes not that clear. 

It is also noted that the completeness is under the selection function of LAMOST. LAMOST survey can only select a small fraction of targets to observe. Therefore, the completeness here means how many OB stars we identify from the LAMOST observed spectra, which is different with the completeness with respect to the intrinsic OB population.

\section{Sub-classification}

To better understand the selected OB stars, assigning spectral sub-type and luminosity class to each of them is critical. We apply MKCLASS, which is an automatic classification package developed by \citet{gra14}, to classify the spectral sub-type and luminosity class for the identified OB stars.

\subsection{MKCLASS}

MKCLASS is designed to classify blue-violet spectra in the MK spectral classification system \citep{kee93}. The direct comparison between the program star and the MK standard stars is carried out by the package during the classification process. The version 1.07 of MKCLASS\footnote{It can be obtained from \url{http://www.appstate.edu/~grayro/mkclass/.}} is used to classify the 101\,086 spectra of 80\,447 objects over the entire \textit{Kepler} field acquired through the LAMOST-\textit{Kepler} project, which was designed to obtain high-quality, low-resolution spectra of a sample of stars in the \textit{Kepler} field with the LAMOST \citep{gra16,cat15}. 

Two MK standard star libraries (i.e., \textit{libnor36} and \textit{libr18}) are distributed with MKCLASS, one based on flux calibrated spectra covered 3800--5600\,\AA\ and with a resolution $R\sim1100$, while the second one based on rectified spectra covered 3800--4600\,\AA\ and with $R\sim2200$. MKCLASS can determine the spectral type in terms of the MK system for a normal star and assess its performance of classification using ``excellent", ``verygood", ``good", ``fair", and ``poor" at the same time. The $\chi^2$ is computed between the program spectrum and the best matched spectrum of the standard star, which is the basis for performance assessment. 
Through applying successively greater amounts of noise to the spectrum of \citet{kur84} solar flux atlas, \citet{gra14} investigated the impact of the $\mathrm{S/N}$ to the quality of MKCLASS outputs. They found that, for the solar spectrum with $\mathrm{S/N}\gtrsim20$, MKCLASS could give the accurate spectral type G2\,V. 

\subsection{Sub-classification of OB stars using MKCLASS}

Since the LAMOST spectra are not flux-calibrated, we use the library \textit{libr18} for the classification. Before performing MKCLASS, we adjust the spectral calibration of the library spectra to make it consistent with the LAMOST spectra. The actual spectral resolution of the LAMOST spectra in blue wavelength is quantified using five strong arc lines at around $4046$, $4358$, $4678$, $4800$, $5090$, $5470$\,\AA\ in the arc spectra~\citep[They are Hg and Cd lines since the blue arc lamp is a HgCd lamp, see][]{han2018}. The median FWHM of these arc lines over $\sim4000$ fibers (a few of dead fibers have been removed) is $2.78\pm0.07$\,\AA.

Therefore, we convolve the library \textit{lib18} to reduce the resolution from 1.8\,\AA\ to 2.8\,\AA\ so that its resolution is consistent with LAMOST spectra. We denote the new modified library as \textit{libr18\_28}. Then we submit the 21\,793 normalized spectra of our identified normal OB stars (sub-dwarfs and white dwarfs are excluded) to MKCLASS. The output results are listed in Table~\ref{tab:OBstars}. The \textit{8th} column is the MK spectral type identified by MKCLASS in the standard format, i.e., the spectral sub-type at first place followed by the luminosity class.

\begin{figure*}
	\plotone{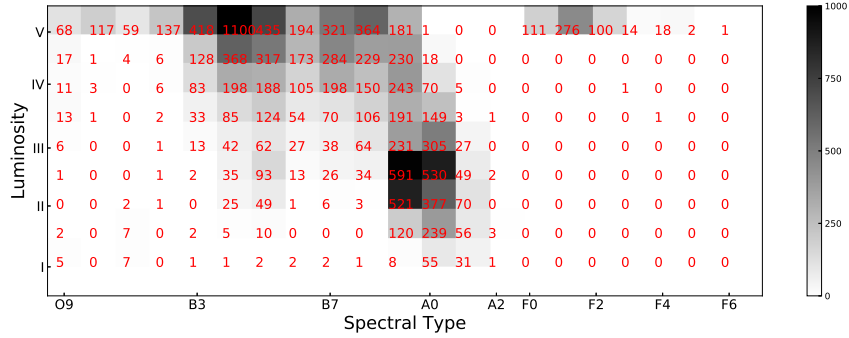}
	\caption{The distribution of stars in spectral sub-type vs. luminosity class plane according to the results from MKCLASS with quality evaluation better than ``good" and including ``good". The number of stars fell in each small bin is drawn in different gray levels and also marked as text in the figure. The rows between two luminosity types are those MKCLASS assigns both types. For instance, the row between II and III are those II-III given by MKCLASS.\\ 
\label{fig:MK-density}}
\end{figure*}

Figure~\ref{fig:MK-density} shows the distribution of stars in spectral sub-type vs. luminosity class plane according to MKCLASS with quality evaluation better than ``good" and including ``good". It shows that most of the OB stars are distributed between B3 and A1 in spectral type. Only 123 are classified as O type stars. 524 stars are even classified as F-type stars. While most of the OB type stars are in luminosity class V (main sequence), two substantial giant/supergiant branches are displayed in the figure: one is located at around B4--B5 and the other is at around B8--A0. If we focus on the overdensities of the distribution, two substantial overdensities can be seen at B4V with 1\,100 samples and B9II-III with 591 stars, respectively. The branches and overdensities reflect the combination of the intrinsic distribution of the OB stars in the spectral and luminosity types and the observational selection effect.  

\begin{figure}
\plotone{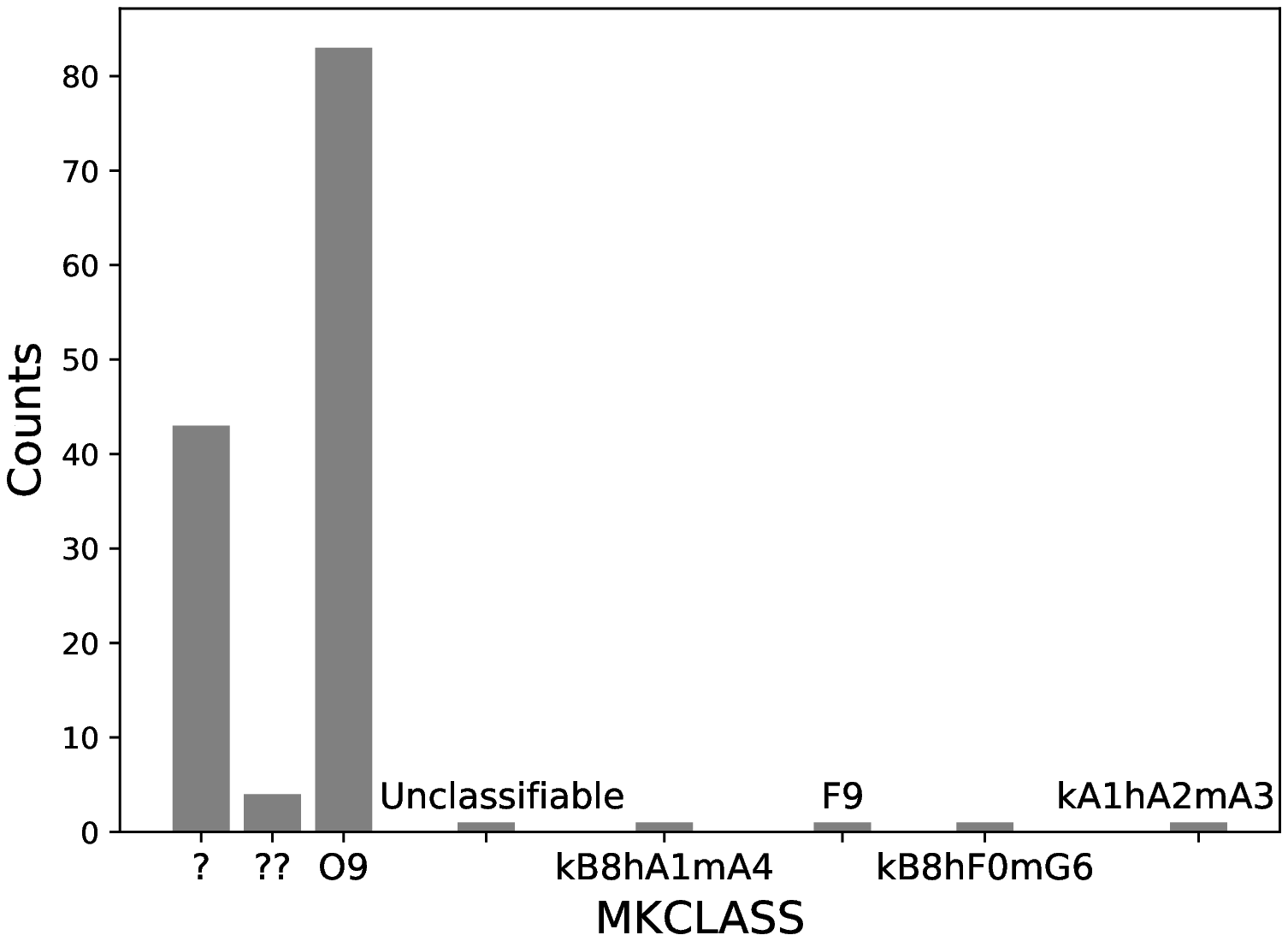}
\caption{Histogram of the results identified by the MKCLASS for 135 spectra of 91 O stars identified in the by-eye analysis.}\label{histo-ostar}	
\end{figure}

According to the online documentation, the libr18 spectral library only contains stars up to spectral type O9\footnote{\url{http://www.appstate.edu/~grayro/mkclass/mkclassdoc.pdf}}. This means that MKCLASS would not be able to identify any star earlier than O9. Figure~\ref{histo-ostar} shows the histograms of the results identified by the MKCLASS for 135 spectra of 91 O stars identified in the by-eye analysis (see Table~\ref{tab:OBstars}). It can be seen that except results with labels ``?" or ``??", most of O stars identified via visual analysis are identified as type O9 by MKCLASS.  

\subsection{Quality flags from MKCLASS}

\begin{figure}
	\plotone{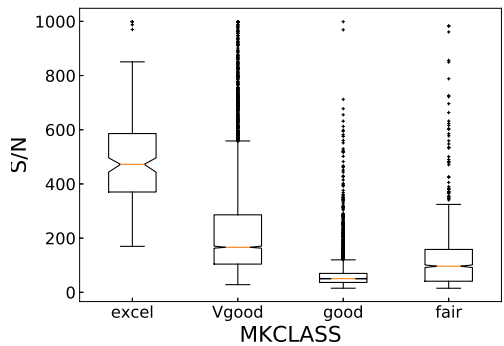}
	\caption{The correlation between the quality flags from MKCLASS and signal-to-nose ratio of the spectra.\label{fig:SNRG}}
\end{figure}

Among the results of sub-classification, 209 spectra are rated as ``excellent'' with the median $\mathrm{S/N}=427$ at \textit{g} band, 8\,109 as ``vgood'' with $\mathrm{S/N}=166$, 3\,279 as ``good'' with $\mathrm{S/N}=50$, 762 as ``fair'' with $\mathrm{S/N}=96$, and 1 as ``poor'' with $\mathrm{S/N}=19$. \citet{gra16} pointed out that spectral types with ``fair'' are uncertain and those with ``poor'' are unreliable. These two quality categories are usually assigned to the spectra with substantial defects or low S/N. 

Figure~\ref{fig:SNRG} shows the correlation between the quality flags from MKCLASS and signal-to-noise ratio of the spectra. It shows that the ``excellent'' results are only for the spectra with signal-to-noise ratio larger than 200. Most of the spectra with signal-to-noise ratio larger than 100 obtain the quality level of ``verygood''. Most of the stars with signal-to-noise ratio lower than 100 are assigned with ``good'' and ``fair''. Note that there are also 273 stars with signal-to-noise ratio larger than 200 assigned with ``good'' and ``fair''. This is mostly because that 1) a few spectra have overestimated signal-to-noise ratio and 2) MKCLASS may significantly misclassifies the sub-types for a few stars. 

\begin{figure}
	\plotone{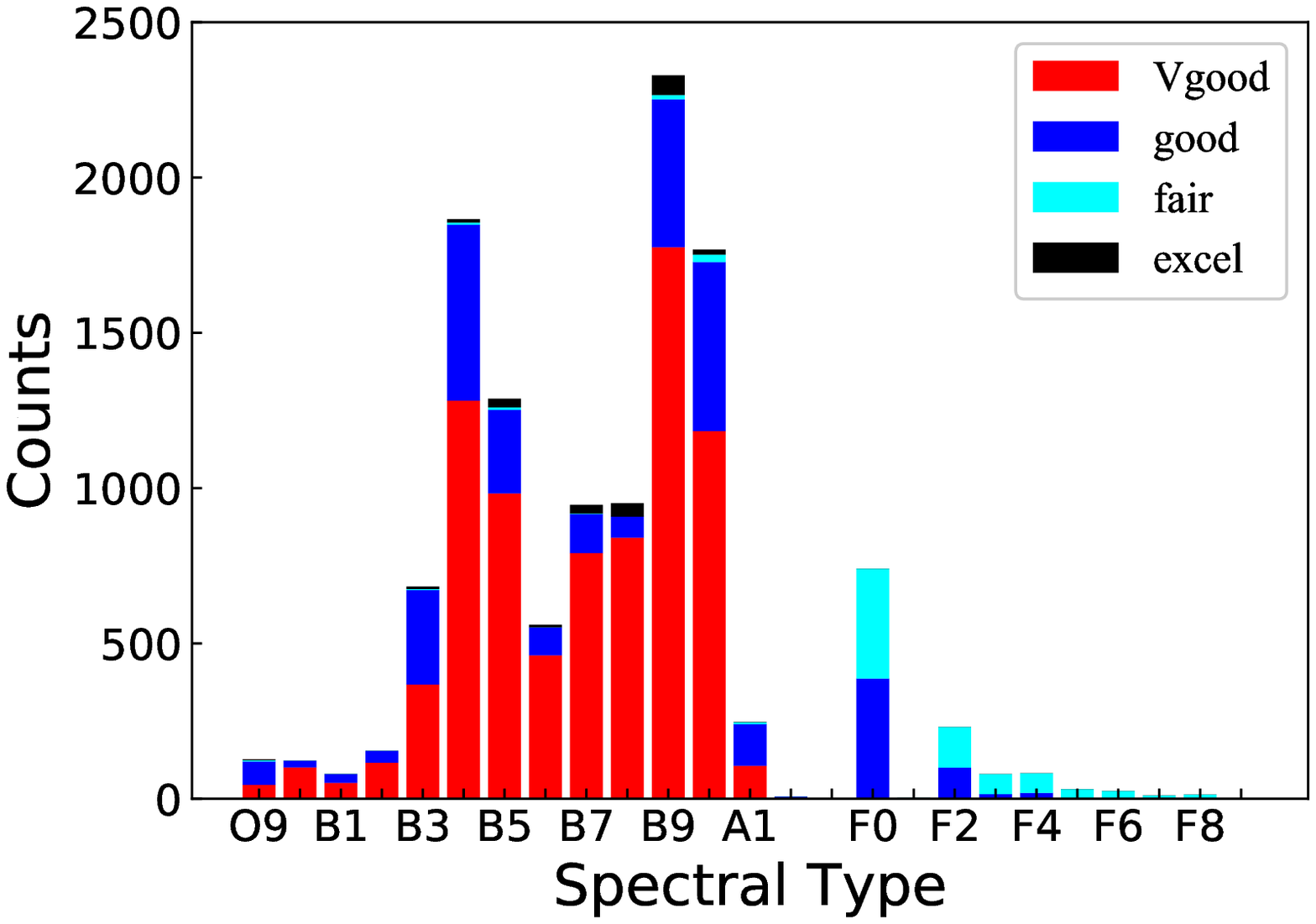}
	\plotone{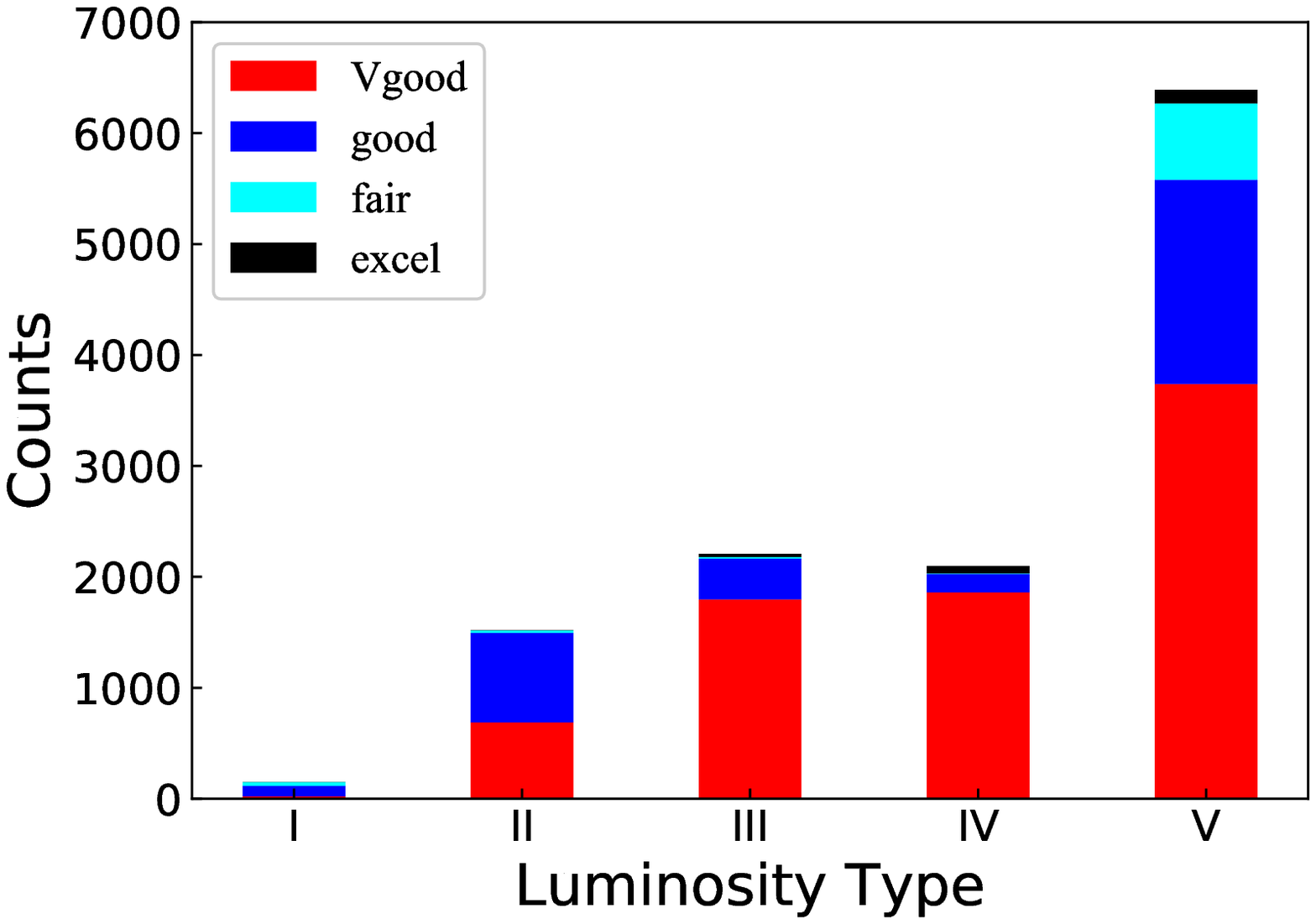}
	\caption{ The distribution of the spectral types (top panel) of the OB stars 
	obtained from MKCLASS. And the bottom panel shows the distribution of the luminosity type. The quality flags are represented by different colors.
\label{fig:spe-lum-MK}}
\end{figure}

The top panel of Figure~\ref{fig:spe-lum-MK} shows the distribution of the spectral types of the OB stars obtained from MKCLASS. It is seen that while not many spectra are qualified as ``excellent'', the majority are the stars with quality of ``verygood'' covering from O9 to A1 type. A trend is that the distribution of the spectral types shows two peaks at B4 and B9, respectively. The two peaks are likely due to the two overdensities located at B4 V and B9 II-III, respectively, displayed in Figure~\ref{fig:MK-density}. 1\,205 OB spectra are misclassified as F type stars with quality level of ``good'' and ``fair''. We analyze some sample of these spectra in section~\ref{sec:MKCLASSissue}. 

The bottom panel shows the distribution of the luminosity class from MKCLASS. As expected, most of the stars are main-sequence (V) stars. The quality flags ``fair'' only appear in luminosity class I and V. 

\subsection{Issues of the MKCLASS results}\label{sec:MKCLASSissue}
Some of the OB spectra are not correctly classified by MKCLASS.

\begin{figure*}
	\plotone{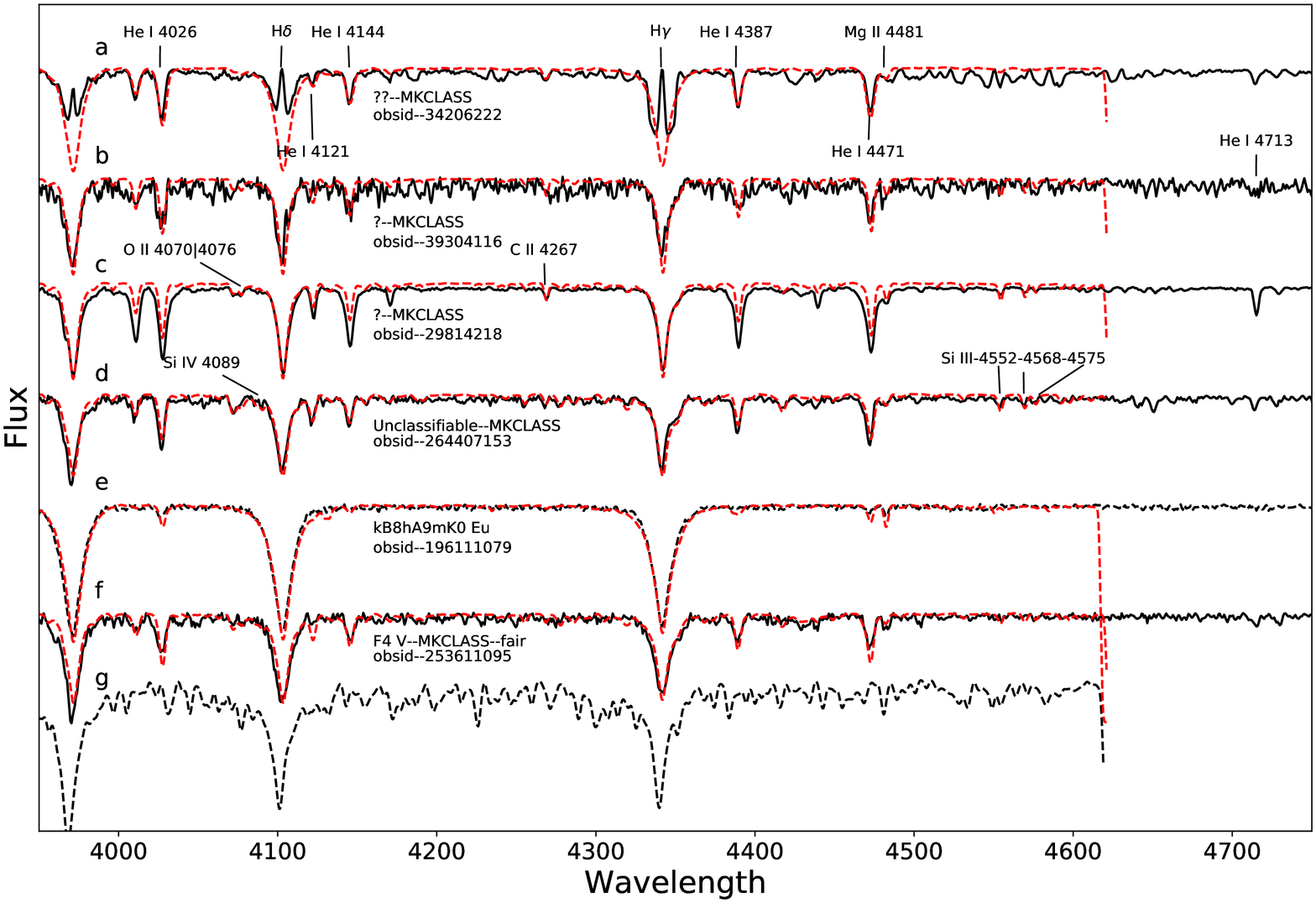}
	\caption{The sample spectra that the MKCLASS assignes labels of question marks. The black solid lines represent the observed spectra, and the red dashed lines represent the proper template spectra manually selected from the library \textit{lib18}, whose resolution have been reduced from 1.8\,\AA\ to 2.8\,\AA. The spectral types from up to down are B2V, B2III, B2III, B1III, B8V, B1V, respectively. A template spectra with the spectra type of F4 V is provided for comparison (black dashed line). \label{fig:MK-question}}
\end{figure*}

MKCLASS assigns question marks to 4\,922 spectra. According to \citet{gra16}, this is likely because that some fluxes in the spectra are negative. However, when we investigate the spectra in details, we find lots of them are normal B type stars. Figure~\ref{fig:MK-question} shows some sample spectra classified by MKCLASS with labels of question marks (see the first to third spectra). From this figure, we can see that this issue is mainly due to the spectra with emission lines (first spectrum), or lower signal to noise (second), or strong helium lines (third).

Meanwhile, it assigns ``Unclassifiable'' for 89 spectra. \citet{gra16} thought that these are because the negative fluxes in the cores of strong lines (e.g. \CaK), which is introduced by the excessive background subtraction. We investigate these spectra and find that many of them look normal (see the fourth spectrum in Figure~\ref{fig:MK-question}). The fourth spectrum seems have stronger He\,I lines than the template spectrum. 

\begin{figure}
\plotone{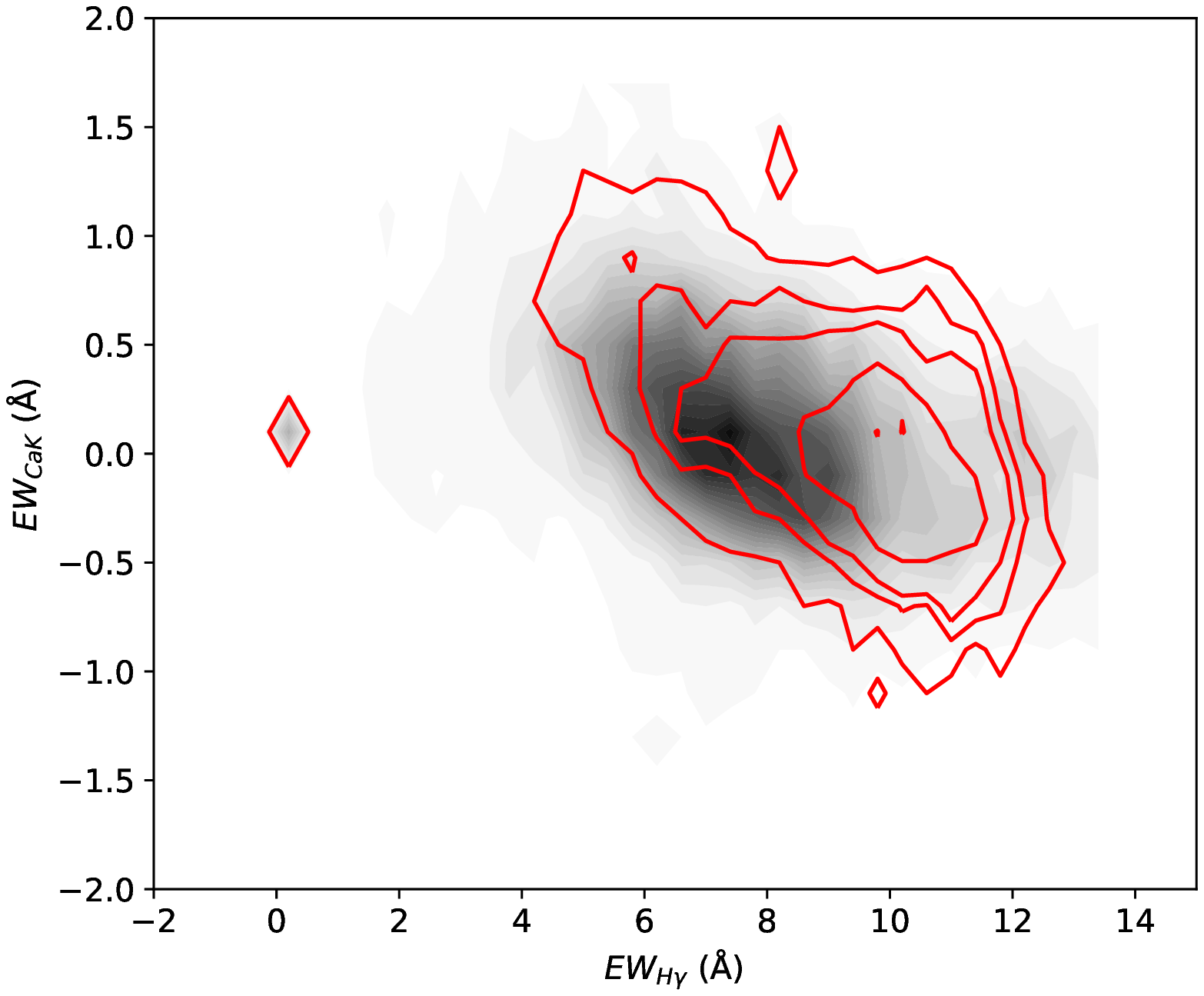}	
\caption{The gray shadow displays the distribution of \CaK\ vs. \Hgamma\ for the stars with successful classifications in MKCLASS. The red contours indicate the similar distribution for the stars with discrepant classifications in MKCLASS.}\label{fig:KH_CaKHg}
\end{figure}

Moreover, MKCLASS cannot converge to a single type but provides discrepant classifications with different criteria for 4\,422 spectra. In these cases, MKCLASS outputs flags like ``kB8hA9mK0 Eu'', which means that the K-line determined spectral type is B8, the hydrogen-line determined type is A9, and the metallic-line clasified type is K0 (as an instance, see the fifth spectrum in Figure~\ref{fig:MK-question}). In the mean time, the star may show strong Eu lines, which should not be true for early type stars. This kind of results are because that MKCLASS determines separate temperature types using the \CaK\ line, the hydrogen lines, and the general metallic lines from the spectra. When these determinations are not consistent, MKCLASS tends to give such a long and complicated special type. There is no quality evaluations given for this kind of classifications. Figure~\ref{fig:KH_CaKHg} shows that the stars with multiple class flags have different distribution of \CaK\ vs. \Hgamma\ against the stars with consistent and normal classification. It is seen that given similar \CaK\, the stars with normal classification have smaller EW of \Hgamma\ than those with inconsistent classes. This is probably because 1) the stars with inconsistent classification may be affected by their fast rotation leading to broader H lines or 2) the stars with inconsistent classification may have lower metallicity \citep{fra10,mar17}.

\begin{figure*}[h]
	\epsscale{0.7}
         \plotone{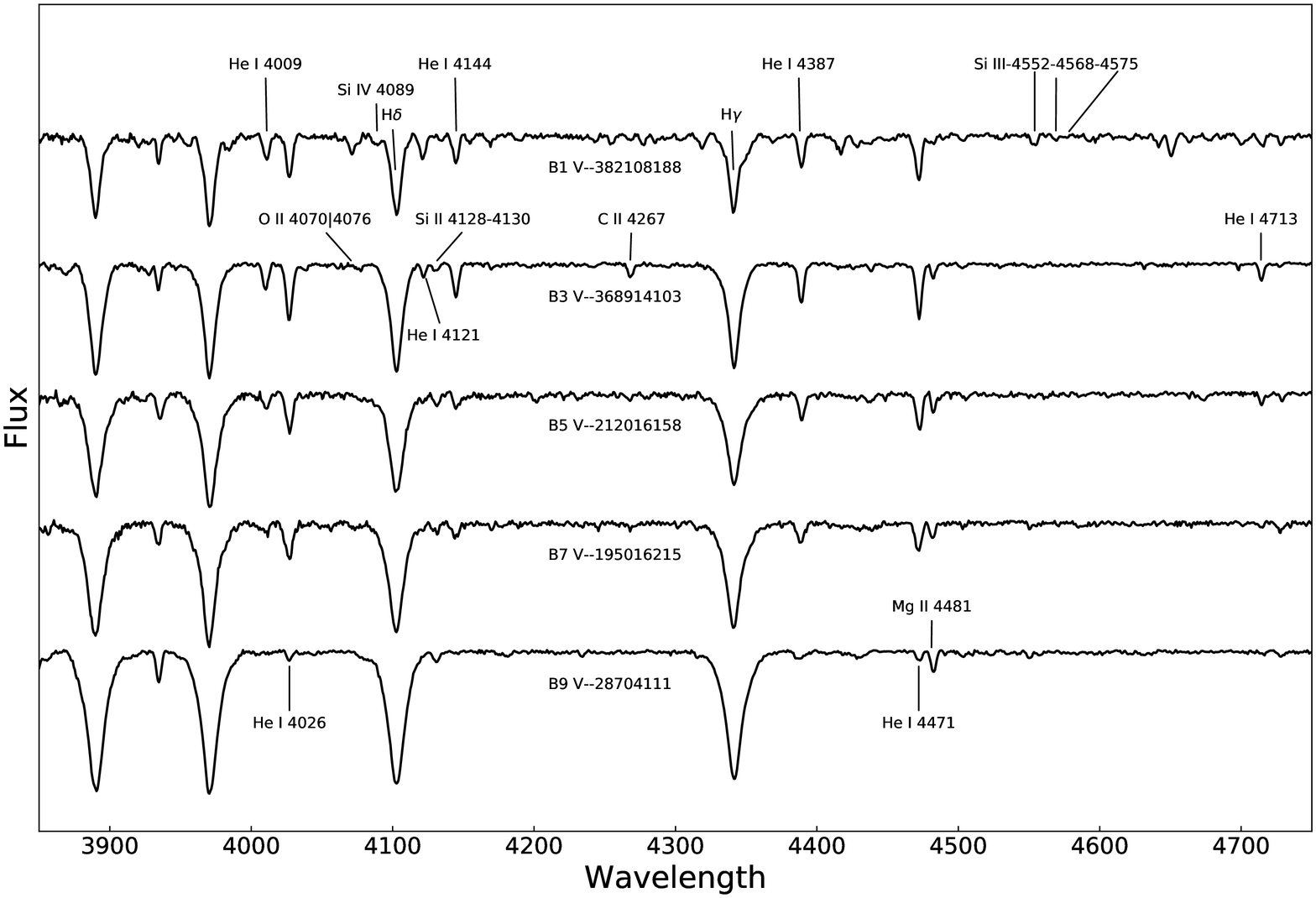}
         \plotone{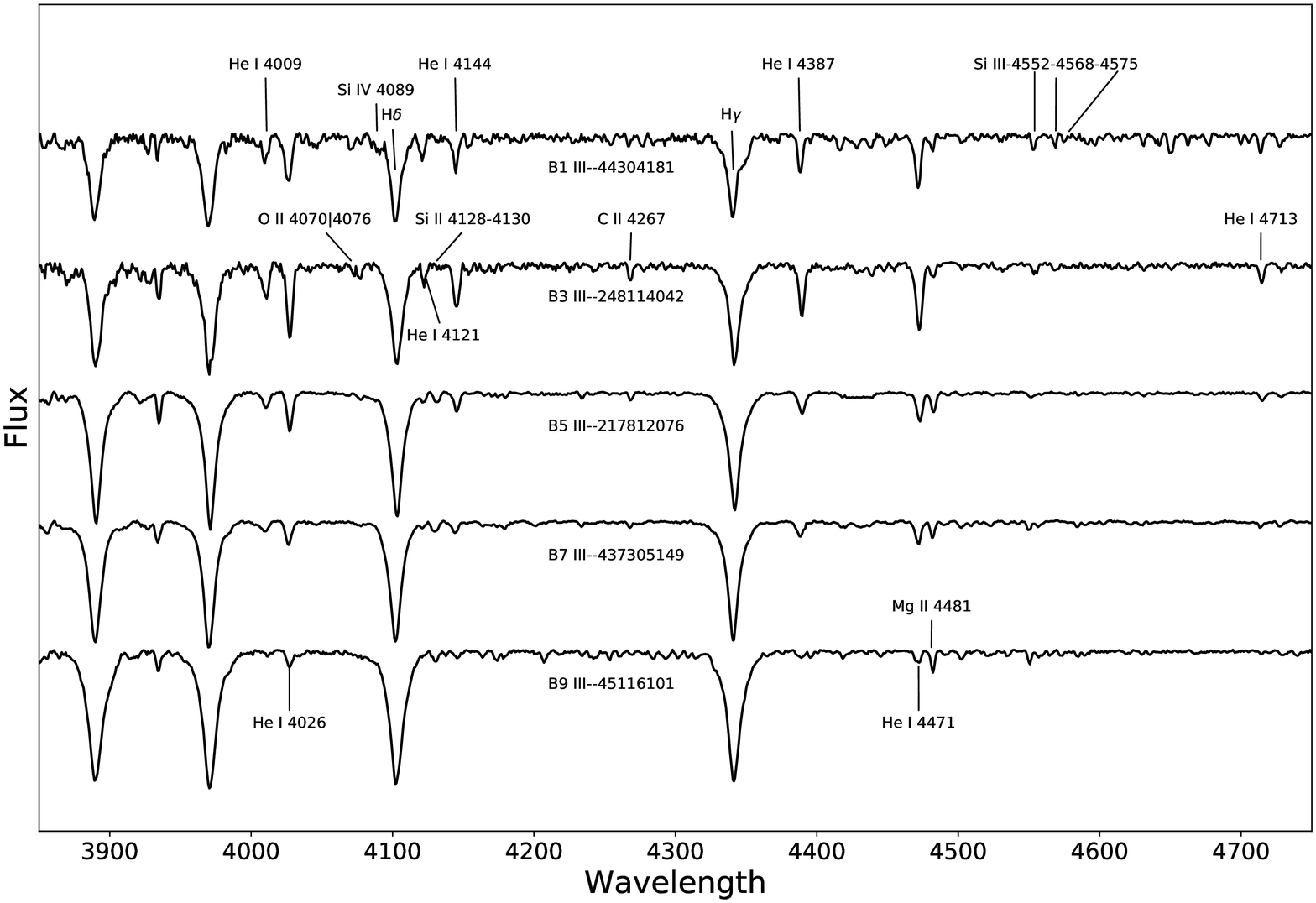}
	\plotone{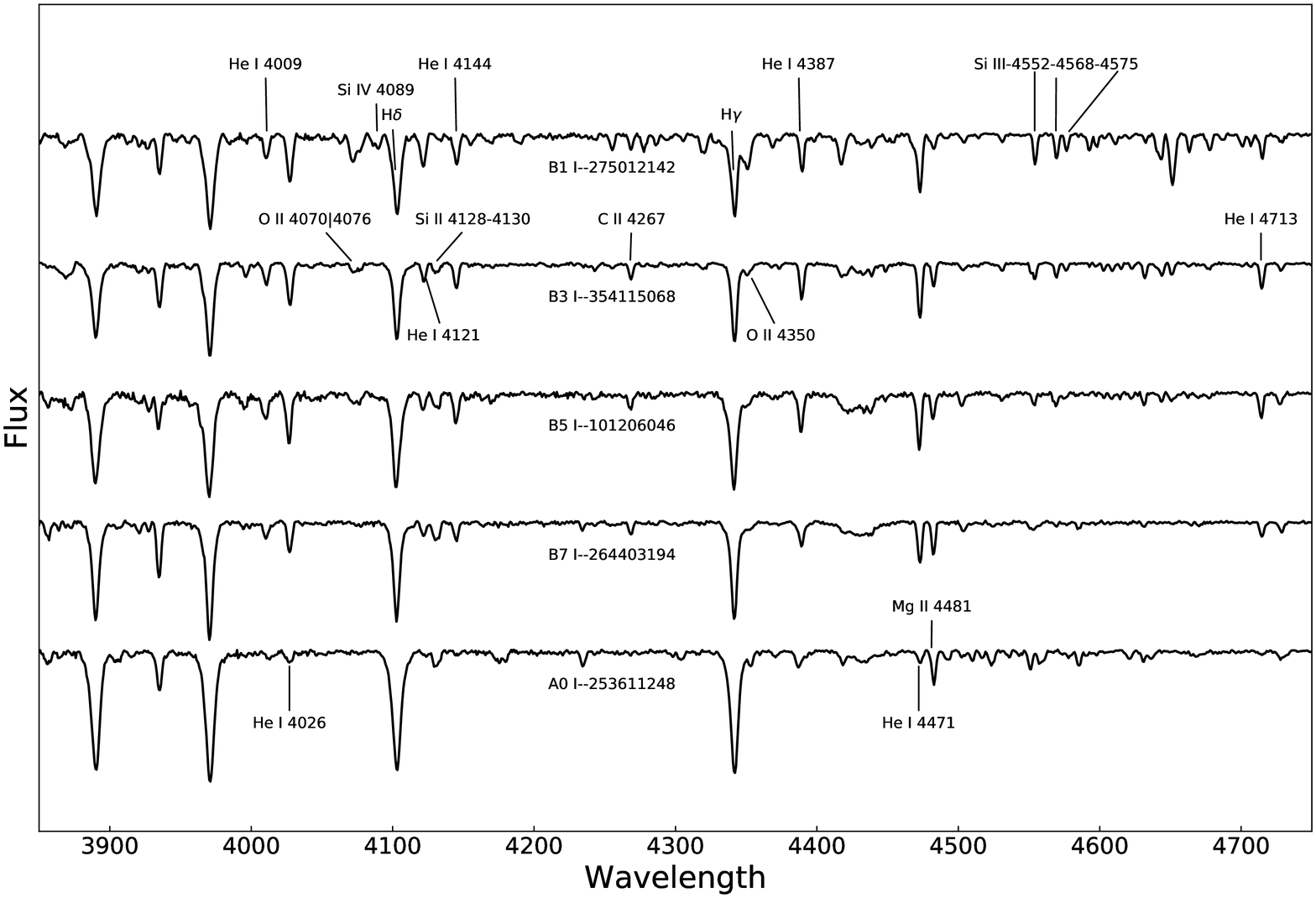}
	\caption{The panels from top to bottom show the sequence of the manually classified luminosity classes, i.e., V, III, I, respectively. And the sequence of manually classified spectral sub-types for sample spectra are displayed in each panel. The corresponding spectral sub-types, luminosity classes and the \obsid\ (the ID of a spectrum in LAMOST catalog) are marked under the corresponding spectra. Important line features are marked in the plots.}\label{fig:luminosity V-I}
\end{figure*}

\subsection{Assess performance of the MKCLASS}
We access the performance of MKCLASS in two ways.

\begin{figure}
	\epsscale{1}
	\plotone{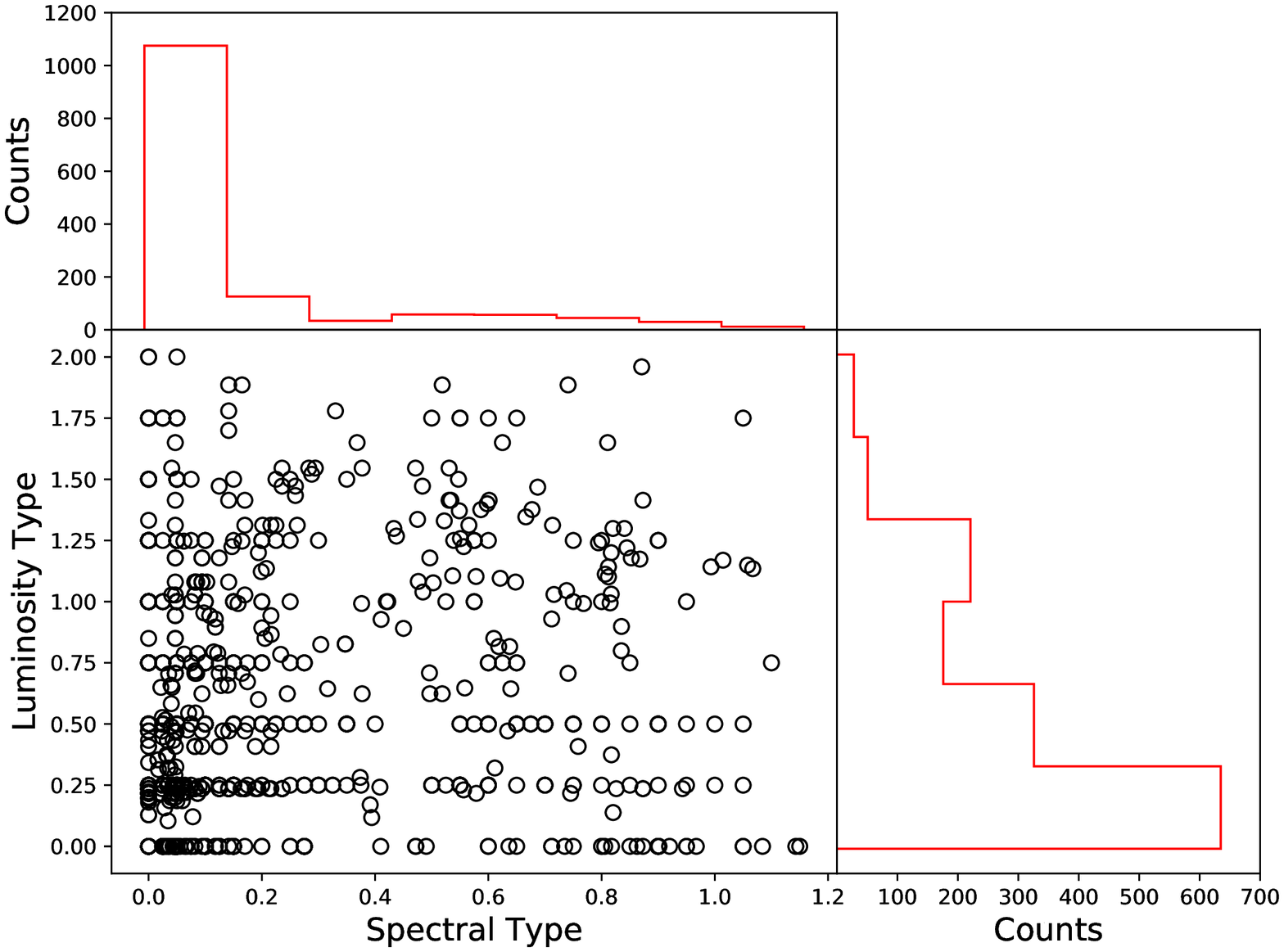}
	\caption{Distribution of the standard deviations of the spectral and luminosity types of the multiply observed  OB stars. The $x$-axis represents the standard deviation of the spectral types classified by MKCLASS in different observations, and the $y$-axis represents the standard deviation for there luminosity types. The numbers in $x$-axis represent for the spectral sub-type such as 0.1-0.9 for O1-O9, 1.0-1.9 for B0-B9. And the numbers in $y$-axis stand for luminosity sub-type, i.e. 1-5 for I-V.
	\label{fig:standard-dev}}
\end{figure}

First we use the OB stars with multiple observations to check the consistency of the MKCLASS. For each observed spectra, we have one MKCLASS classification. For the OB stars with multiple observations, we can compare whether MKCLASS assigns consistent classes for same stars based on different epoch spectra. Figure~\ref{fig:standard-dev} shows the distribution of the standard deviations of the spectral and luminosity types of the repeatedly observed OB stars. We can see that for most of the stars with multiple observations, the MKCLASS can give a consistent spectral and luminosity types with dispersions of about 1 sub-type and 0.5 luminosity level, respectively.

Second, we arbitrarily select 1\,000 OB stars to independently classify them by eyes. When we do the manual inspection, we do not know the results of classification from MKCLASS. In other word, this test is a blind test. 

Based on \citet{gra09} and considering the low-resolution characteristics of the LAMOST spectra, we select a set of the most prominent features to discriminate the spectral and luminosity types. 

For O-type stars, the prominent features include \HeI\ and \HeII\ lines at 4026, 4200, 4541, and 4686\,\AA, a group of \NIII\ lines located at 4634--4640--4642\,\AA. \HeII\,4200, 4541 and \HeI/II\,4026 reach their maximum at O5. They are weakened when the luminosity type changes from dwarfs to giants. They become very weak or even become emission lines in the spectra of supergiants. The \NIII\ lines strengthen from O2 to O4 and then weaken when the spectral type changes to O7. Meanwhile, these lines also become stronger in giant stars \citep{walborn1990,lennon1997,gra09,sot11}.

For B-type stars, the line ratio between \HeI\,4471 and \MgII\,4481\,\AA\ is the well known indicator of the spectral type. Luminosity types of B-type stars are usually discriminated by looking at the ionized metal lines, e.g. \SiII\,4128--4130, \SiIII\,4552--4568--4575, \CII\,4267, \NII\,3995\,\AA\ etc.

More detailed criteria for discrimination of the spectral sub-type and luminosity class are listed in Table~\ref{tab:subclass}. In this table, the first row of each spectral sub-type gives the general features of the sub-type. The second row of each sub-type provides the criteria of luminosity classification.

Based on all these criteria we obtain the spectral sub-types and luminosity classes of all the arbitrarily selected OB stars. Compared to MKCLASS results, 646 of the 1\,000 spectra are assigned meaningful sub-classes (not ``?'' or ``Unclassifible'') with quality flag of at least ``good'' by MKCLASS. We then assess the performance of MKCLASS only using these 646 ``good'' or better spectra.

\begin{table*}
	\begin{center}
        \caption{Criteria for classifying OB stars in the temperature dimension.}\label{tab:subclass}
        \begin{tabular}{|c|c|c|c|}
      \hline
      \multirow{2}{*}{\diagbox[width=6em, trim=r]{SpType}{LumType} }& \multicolumn{3}{ c |}{Criteria} \\
       \cline{2-4}
      & V & III & I\\
      \hline
       \multirow{3}{*}{O2-O3} & \multicolumn{3}{l |}{$\rm{He\,I+II}\,\lambda4026<\rm{He\,II}\,\lambda4200$; weak N III 4634-40-42 emission\tablenotemark{a}  $\uparrow$\tablenotemark{b}; 
       weak Si\,IV\,4089-4116 emission $\uparrow$;} \\
       &  \multicolumn{3}{ l |}{median weak N\,IV\,4058 emission $\uparrow$; weak N\,V\,4604-20 $\downarrow$} \\
       \cline{2-4}
      & weak He\,II 4686 & weaker He\,II 4686 than class V  & weak He\,II 4686 emission \\
       \hline
        \multirow{4}{*}{O4} & \multicolumn{3}{l |}{$\rm{He\,I+II}\,\lambda4026<\rm{He\,II}\,\lambda4200$; weak N III 4634-40-42 emission $\uparrow$; 
       weak Si\,IV\,4089-4116 emission $\uparrow$;} \\
       &  \multicolumn{3}{ l |}{weak N\,V\,4604-20 $\downarrow$} \\
       \cline{2-4}
      & weak He\,II 4686 & weaker He\,II 4686 than class V  & weak He\,II 4686 emission; \\
      & weak He\,I\,4471 & & N\,IV\,4058 emission \\
       \hline
        \multirow{4}{*}{O5} & \multicolumn{3}{l |}{$\rm{He\,II}\,\lambda4541>\rm{He\,I}\,\lambda4471$; N III 4634-40-42 emission $\uparrow$ } \\
       \cline{2-4}
      & $\rm{He\,II}\,\lambda4686\approx\rm{He\,II}\,\lambda4541$ & $\rm{He\,II}\,\lambda4686<\rm{He\,II}\,\lambda4541$  & N\,III\,$\lambda$4634-40-42\,$\approx\rm{He\,II}\,\lambda4686$ \\
      & He\,II\,4686 & weak He\,II 4686 & He\,II 4686 emission \\
      & &weak Si\,IV 4116 emission & Si\,IV 4116 emission \\
       \hline
       \multirow{4}{*}{O6} & \multicolumn{3}{l |}{$\rm{He\,I+II}\,\lambda4026\approx\rm{He\,II}\,\lambda4200$; N III 4634-40-42 emission $\uparrow$; Si\,IV\,4089-4116 $\uparrow$ } \\
       \cline{2-4}
      & $\rm{He\,II}\,\lambda4541>\rm{He\,I}\,\lambda4471$ & $\rm{He\,II}\,\lambda4541\approx\rm{He\,I}\,\lambda4471$  & $\rm{He\,II}\,\lambda4541<\rm{He\,I}\,\lambda4471$ \\
      & median weak He\,II\,4686 & weak He\,II 4686 & He\,II 4686 emission; weak N\,III\,$\lambda4097$ \\
       \hline
       \multirow{3}{*}{O7} & \multicolumn{3}{l |}{$\rm{He\,I}\,\lambda4471\approx\rm{He\,II}\,\lambda4541$; N III 4634-40-42 emission $\uparrow$; weak Si\,IV\,4089-4116 $\uparrow$ } \\
       \cline{2-4}
      & $\rm{He\,II}\,\lambda4686>\rm{He\,I}\,\lambda4471$ & $\rm{He\,II}\,\lambda4686\approx\rm{He\,I}\,\lambda4471$  & Si IV\,4686-4504 emission \\
      &  &  & He\,II 4686 emission \\
       \hline
       \multirow{5}{*}{O8-O9} & \multicolumn{3}{l |}{$\rm{He\,I}\,\lambda4471>\rm{He\,II}\,\lambda4541$; weak Si\,IV\,4089 $\uparrow$ } \\
       \cline{2-4}
      & $\rm{C\,III}\,\lambda4650>\rm{He\,II}\,\lambda4686$ &$\rm{C\,III}\,\lambda4650>\rm{He\,II}\,\lambda4686$ & Si IV\,4686-4504 emission \\
      & $\rm{He\,II}\,\lambda4686>\rm{He\,I}\,\lambda4713$  & $\rm{He\,II}\,\lambda4686>\rm{He\,I}\,\lambda4713$ & He\,II 4686 emission \\
       & $\rm{Si\,IV}\,\lambda4089<\rm{C\,III}\,\lambda4187$  & $\rm{Si\,IV}\,\lambda4089\approx\rm{C\,III}\,\lambda4187$ & weak N\,III 4097 \\
        & & weak Si IV\,4686-4504 emission & N\,III 4634-40-42 emission \\
       \hline
       \multirow{4}{*}{B0} & \multicolumn{3}{l |}{Si III 4552-68-75; weak He II 4200; O II 4070-76 $\uparrow$; Si\,IV\,4089 $\uparrow$; He II 4686$\downarrow$ } \\
       \cline{2-4}
      &  $\rm{He\,II}\,\lambda4686>\rm{He\,I}\,\lambda4711$& $\rm{He\,II}\,\lambda4686<\rm{He\,I}\,\lambda4713$ & $\rm{Si\,IV}\,\lambda4089>\rm{He\,I}\,\lambda4121$ \\
       & weak O II 4350  & O II 4350, N II 3995 & strong O II 4350; N\,III 4097; N II 3995  \\
       \hline
       \multirow{4}{*}{B1} & \multicolumn{3}{l |}{Si III 4552-68-75 $\uparrow$; N\,II 3995 $\uparrow$; O II 4070-76 $\uparrow$; Si\,IV\,4089 $\uparrow$} \\
       \cline{2-4}
      &  $\rm{Si\,III}\,\lambda4552\approx\rm{Si\,III}\,\lambda4568$& $\rm{Si\,III}\,\lambda4552\approx\rm{Si\,III}\,\lambda4568$ & $\rm{Si\,III}\,\lambda4552>\rm{Si\,III}\,\lambda4568$ \\
       & $\rm{Si\,III}\,\lambda4568\approx\rm{Si\,III}\,\lambda4575$  &  $\rm{Si\,III}\,\lambda4568\approx\rm{Si\,III}\,\lambda4575$ &  $\rm{Si\,III}\,\lambda4568>\rm{Si\,III}\,\lambda4575$  \\
        & & weak O II 4350 & strong O II 4350  \\
       \hline
       \multirow{3}{*}{B2-B3} & \multicolumn{3}{l |}{ $\rm{Si\,I}\,\lambda$4128-30$<\rm{He\,I}\,\lambda4121$; $\rm{He\,I}\,\lambda4471>\rm{Mg\,II}\,\lambda4481$; weak Si III 4552-68-75 $\uparrow$; C\,II 4267; O II 4070-76 $\uparrow$}\\
       \cline{2-4}
      &  weak He I 4121 & N II 3995 & weak O II 4350, N II 3995 \\
       \hline
       \multirow{4}{*}{B5} & \multicolumn{3}{l |}{$\rm{He\,I}\,\lambda4471>\rm{Mg\,II}\,\lambda4481$; C\,II 4267; Si II 4128-30 }\\
       \cline{2-4}
      &  Si III 4552-68-75 absent & weak Si III 4552-68-75 & $\rm{Si\,I}\,\lambda$4128-30$<\rm{He\,I}\,\lambda4121$ \\
       &   & weak Si III 4552-68-75 & median weak Si III 4552-68-75; N II 3995 \\
       \hline
       \multirow{3}{*}{B7} & \multicolumn{3}{l |}{$\rm{He\,I}\,\lambda4471>\rm{Mg\,II}\,\lambda4481$; C\,II 4267; weak Fe II 4233}\\
       \cline{2-4}
      &  He I 4121 absent & $\rm{Si\,I}\,\lambda$4128-30$\approx\rm{He\,I}\,\lambda4121$& $\rm{Si\,I}\,\lambda$4128-30$<\rm{He\,I}\,\lambda4121$; weak N II 3995 \\
       \hline
        \multirow{4}{*}{B8} & \multicolumn{3}{l |}{weak Fe II 4233; weak Si II 4128-30 $\uparrow$}\\
       \cline{2-4}
      &  $\rm{He\,I}\,\lambda4471<\rm{Mg\,II}\,\lambda4481$ & $\rm{He\,I}\,\lambda4471<\rm{Mg\,II}\,\lambda4481$ & $\rm{He\,I}\,\lambda4471\approx\rm{Mg\,II}\,\lambda4481$ \\
        &   &  & $\rm{Si\,I}\,\lambda$4128-30$>\rm{He\,I}\,\lambda4144$;  C II 4267 \\
       \hline      
        \multirow{2}{*}{B9} & \multicolumn{3}{l |}{$\rm{He\,I}\,\lambda4471<\rm{Mg\,II}\,\lambda4481$; weak Si II 4128-30 $\uparrow$; weak Fe II 4233}\\
       \cline{2-4}
        & weak He I 4026   & $\rm{Si\,II}\,\lambda$4128-30$\approx\rm{He\,I}\,\lambda4026$ & $\rm{Si\,II}\,\lambda$4128-30$>\rm{He\,I}\,\lambda4026$ \\
       \hline 
        \end{tabular}       
        \end{center}
        	\tablenotetext{a}{This represents emission line, and others are absorption lines.}	
	\tablenotetext{b}{This means the lines strengthen with luminosity class from V to I.}
\end{table*}

To validate the manual classification, we select the representative spectra and sort them in the sequence of the manually classified luminosity classes in Figure~\ref{fig:luminosity V-I}. And the sequence of manually classified spectral sub-types for sample spectra are displayed in each panel. 

In the top panel, in which the sequence of spectral sub-types for main-sequence stars are displayed, the features that indicate the change of the spectral sub-types are clearly seen. For instance, \HeI\,4471 are firstly strengthened and then weakened from B1 V (top) to B9 V (bottom). And the most insensitive \HeI\,4471 displays at B3 V, while \MgII\ gradually strengthens when the spectral types become later. The other \HeI\ lines show similar trend to 4471\,\AA\ line. In the mean time, the Balmer lines become stronger from top to bottom.

The bottom panel shows the sequence of spectral sub-types for supergiants. It is seen that essentially the metal lines such as \SiIII, \SiIV, \OII\ and \CII\ are stronger in supergiant, while Balmer lines are shallower.

The variance of the spectral line features along either the sequence of spectral sub-types or luminosity classes are consistent with \citet{gra09}. This confirms that the manually inspection and classification of the subsamples of the OB stars is reliable.

\begin{figure}
	\plotone{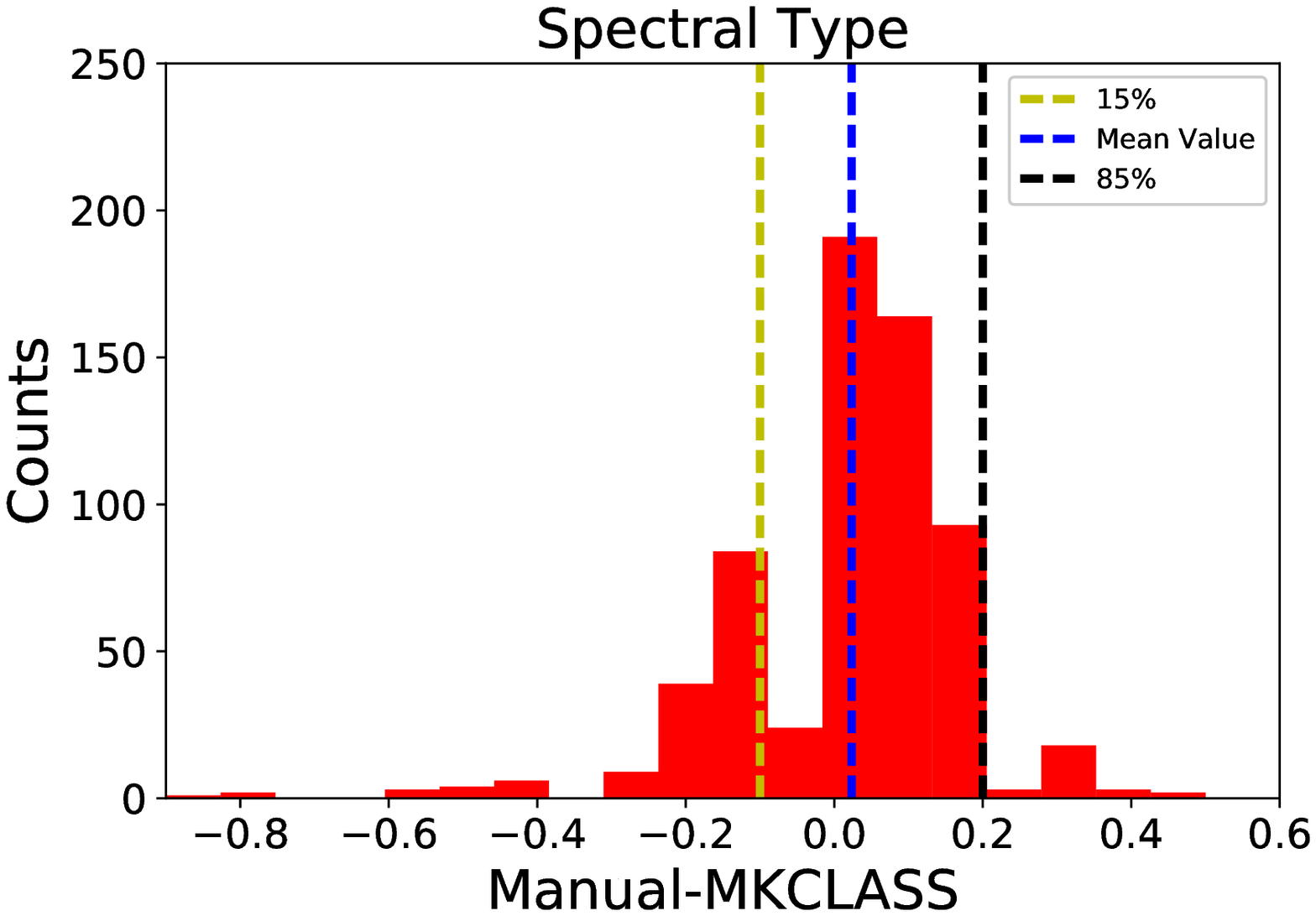}
	\plotone{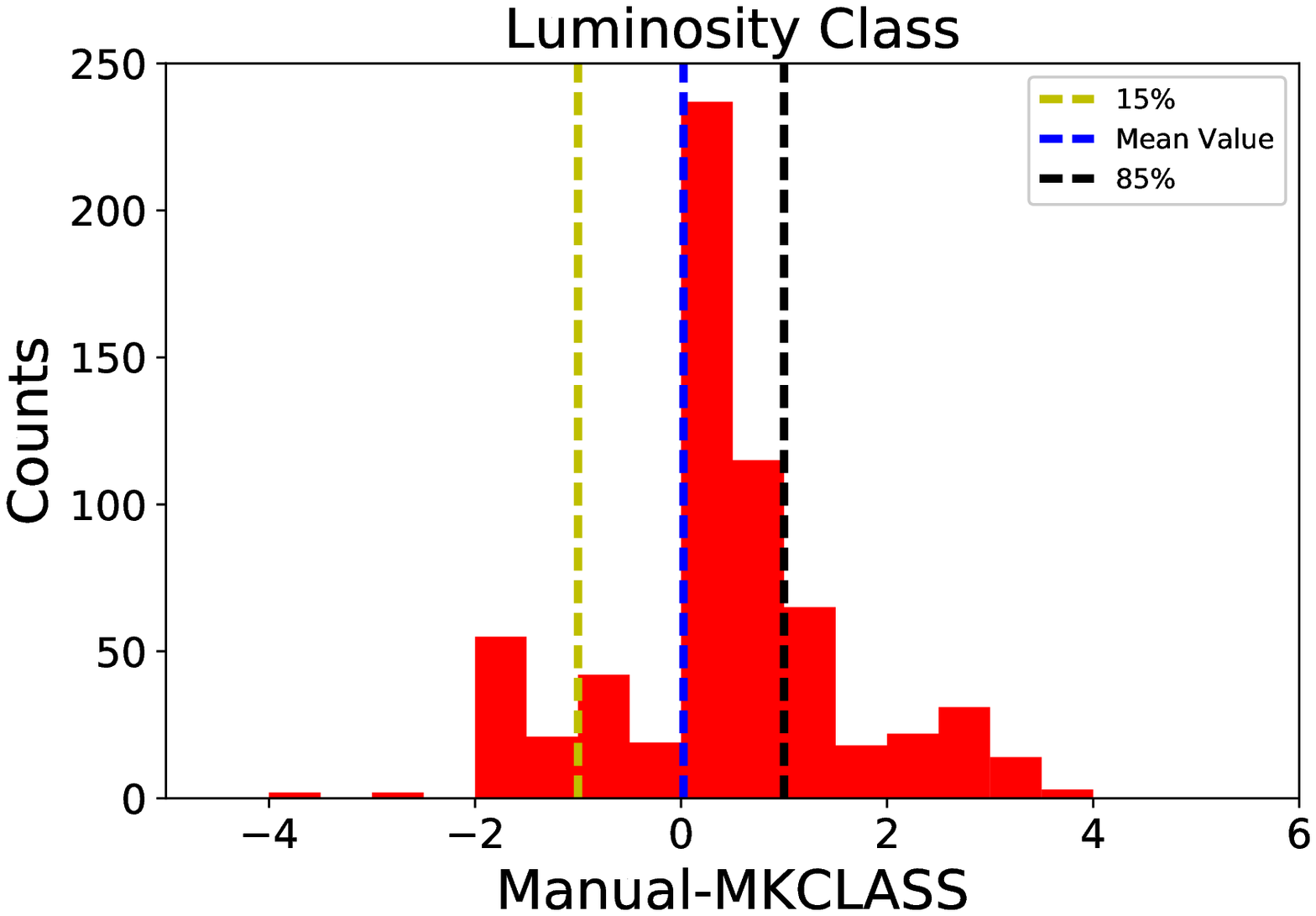}
	\caption{The top panels shows the histogram of the difference of the spectral sub-type between the manual and MKCLASS classified results, i.e. MKCLASS$-$Manual class. The bottom panel shows the histogram of the difference of the luminosity class. The blue dashed lines are the mean values of the difference. The green and black dashed lines indicate the 15\% and 85\% percentile, respectively.}\label{fig:spe-TW-MK}
\end{figure}

We compare the spectral sub-types and luminosity sub-types between the manually classification results and those from MKCLASS. Figure~\ref{fig:spe-TW-MK} shows the histograms of the difference (Manual class-MKCLASS) of the spectral sub-types (top panel) and luminosity classes (bottom panel). For convenience,  the numbers are used to represent the spectral sub-type and luminosity sub-type, which are 0.1-0.9 for O1-O9, 1.0-1.9 for B0-B9, and 1-5 for I-V, respectively. 

\begin{figure*}
	\epsscale{0.85}
	\plotone{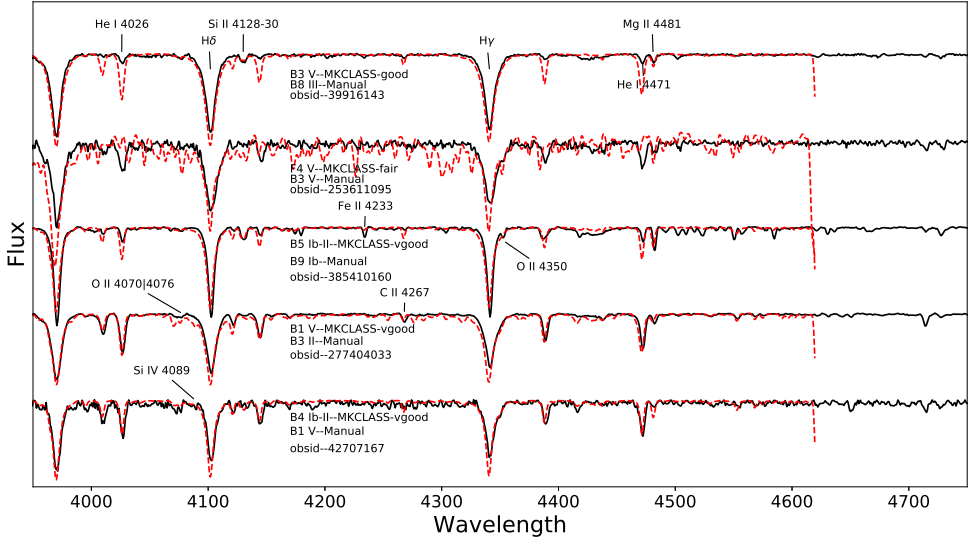}
	\caption{The sample spectra that have substantially inconsistent classifications between manually inspection and MKCLASS. The best-fit spectra from MKCLASS library are also provided with red dashed lines for comparison. \label{fig:MK-different}}
\end{figure*}

We find that, in the top panel, the mean value of the difference (blue dashed line) is at around $+0.025$, implying that the manual classified spectral sub-type is consistent with the results from MKCLASS. Moreover, the 15\% and 85\% percentiles in the differences show that the dispersion between the two methods is less than 1 sub-type. This confirms that, statistically, MKCLASS can obtain quite reliable results for most of the OB type stars. 

 The difference of luminosity class shown in the bottom panel seems more scattered than the spectral sub-type, although the mean value of the difference of the luminosity class is also located at around zero. This is reasonable that the features sensitive to luminosity are mostly weak and are not easy to be well investigated in low-resolution spectra. The 15\% and 85\% percentiles show that the dispersion between the two methods is about one level of luminosity class. 

Note from Figure~\ref{fig:spe-TW-MK} that there are 28 and 52 stars show larger than $3\sigma$ difference in the spectral and luminosity type between the manual approach and MKCLASS. We double-checked these spectra and find that mostly MKCLASS has mistakenly assigned sub-class. Figure~\ref{fig:MK-different} shows some samples. The first spectrum mis-classified as ``B3 V'' in MKCLASS is actually a ``B8 III'' type star according to its weaker He I lines, similar strength between He I 4471 and Mg II 4481\,\AA, and visible weak Si II 4128-30 lines. The second spectrum, which is obviously a ``B3 V'' star according to there relative stronger \HeI\ 4471 than \MgII\ 4481\,\AA, is misclassified as ``F4 V'' by MKCLASS. 
MKCLASS assigns a type of ``B5 Ib-II'' to the third spectrum, which is actually a ``B9 Ib'' type star with its weaker He I 4471 and middle strong Mg II 4481\,\AA. The fourth spectrum should be ``B3 II" type star with its weaker Balmer lines and weak O II 4070/4076\,\AA, which is mis-classified as `` B1 V". MKCLASS assigns ``B4 Ib-II'' to the fifth spectrum, which should be ``B1 V'' based on its visible Si IV 4089 and absence of C II 4267.

Nevertheless, statistically, MKCLASS performs pretty well for most of the OB type stars. Misclassifications in a small fraction of samples may be due to the lack of the template OB type spectra in MKCLASS. Indeed, the template spectra are mostly from the stars nearby the Sun, while the LAMOST OB samples distribute in a wide range of Galactocentric radii in the Galactic outer disk. As a consequence, the LAMOST OB stars are in general have lower matellicity than the template providing a radial metallicity gradient~\citep{daflon2004}. Therefore, given the same spectral sub-type (which is a proxy of effective temperature) and luminosity class (a proxy of surface gravity), the corresponding weak metal lines for the observed samples are not necessarily be same as the template spectra. Finally, relatively larger rotation of the early type stars may be another reason to distort the classification.

\section{Discussion and conclusions}

\begin{figure}
        \includegraphics[scale=0.5]{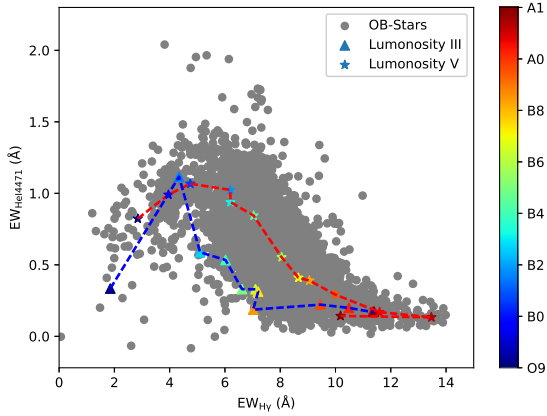}
	\caption{Distribution of the OB type stars in \HeI\ (4471\,\AA) vs. \Hgamma\ (4340\,\AA) plane (filled circles). The colorful starisks indicate the median positions of the spectral sub-types coded with colors for luminosity class V stars. The triangles indicate the locus of the luminosity class III stars with the color-coded spectral sub-types. \label{fig:he-hdelta}}
\end{figure}

\subsection{Distribution in line indices space}
Figure~\ref{fig:he-hdelta} shows the distributions of the identified OB stars in \HeI\ (4471\,\AA) vs. \Hgamma\ (4340\,\AA) plane. It can be seen that the MKCLASS provided spectral sub-types for V and III stars are well arranged as sequences. The later type stars (reddish) are located at the bottom-right corner with larger EWs of \Hgamma\ and almost no \HeI\ lines, while the earlier type stars (bluish) are located the bottom-left corner with both weaker \Hgamma\ and \HeI\ lines. The ``B2 V'' and ``B3 III'' stars show strongest EW of \HeI\ among all the spectral sub-types. This is roughly consistent with \citet{gra09}. 

Although the B giants and dwarfs have similar trends in the \HeI\ (4471\,\AA) vs. \Hgamma\ plane, the locus of the giant stars is roughly located below that of the dwarf stars. This means that given same strength of the H lines, the giant stars have weaker \HeI\ lines than the dwarf stars.

It is also noted that a dozen of stars show stronger \HeI\ than the majority and thus form a branch at the top of the figure. They should be the He-enhanced stars.

\subsection{Spatial distribution}

\begin{figure}
	\plotone{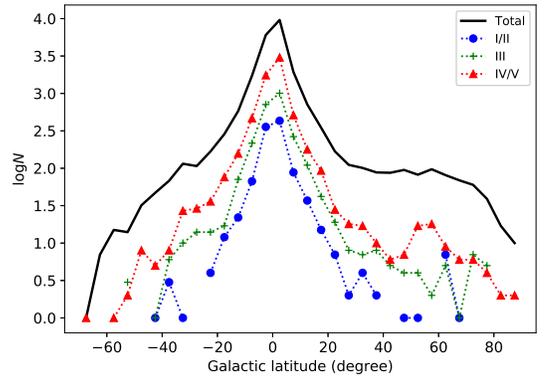}
       \caption{The distributions of OB stars in Galactic latitude. The black line indicates the distribution of all OB stars, while the red triangles, green crosses, and blue filled circles display the distributions for the IV/V, III, and I/II type stars. The $y$-axis is in logarithmic scale.\label{fig:spatial}}
\end{figure}

\begin{figure*}[htp!]
	\plotone{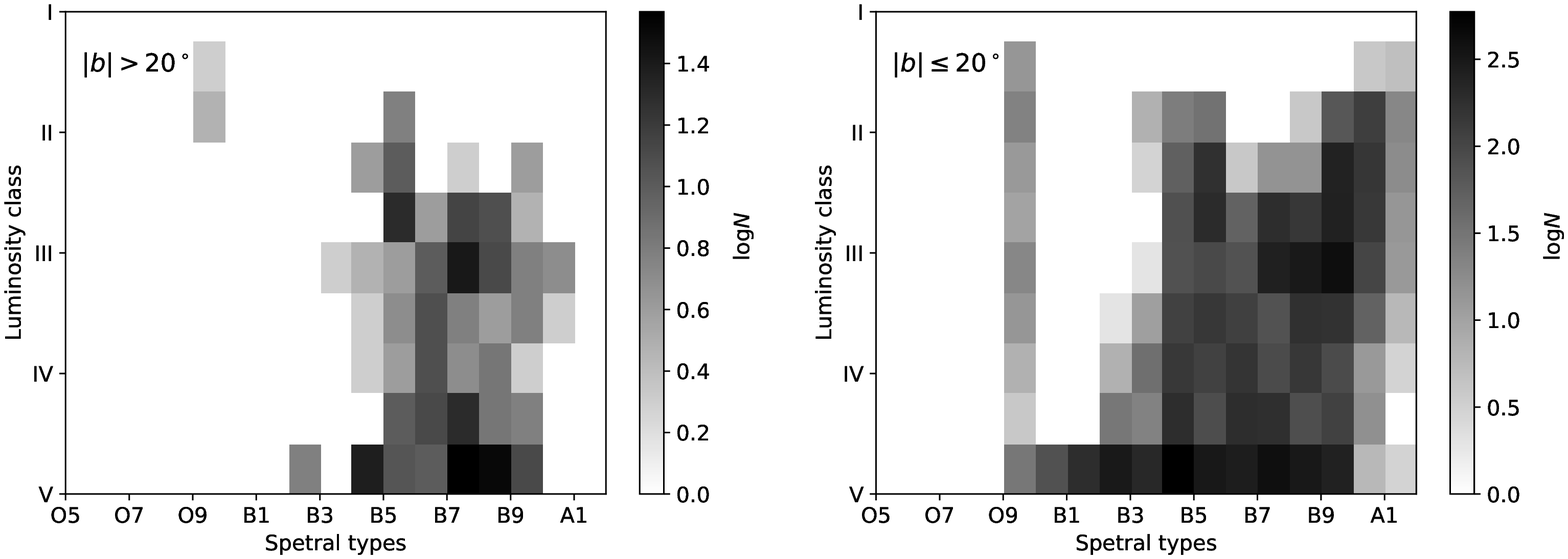}
	\caption{The left panel shows the distribution of the identified OB stars with Galactic latitude larger than 20$^\circ$ in luminosity class vs. spectral sub-type plane. The gray level indicates the number of stars at each bin in logarithmic scale. The right panel is similar but for the OB stars with Galactic latitude smaller than 20$^\circ$.}\label{fig:class_spatial}
\end{figure*}

Figure~\ref{fig:spatial} shows the distribution of all 21\,793 OB spectra in Galactic latitude. While most of them are located in the low Galactic latitudes as expected, 886 of them located at high Galactic latitude ($>20^\circ$ or $<-20^\circ$), in which few massive stars should be found. Figure~\ref{fig:class_spatial} shows that the distributions of the OB stars in Galactic latitude larger than 20$^\circ$ (left panel) and smaller than 20$^\circ$ (right panel) are substantially different, which may either due to the observational selection effect or hint a different origin of the high Galactic latitude OB stars. The stars located in high Galactic latitude are mostly concentrated between B3 and A0, while the stars located at low Galactic latitude shows more spectral types earlier than B3 and also a substantial O9-B0 branch. 

A few channels can explain the high Galactic latitude OB stars. First, some of them are scattered massive stars originally formed in the disk \citep{martin2004}. Second, some of them may be formed \emph{in-situ} in the halo \citep{kee86}. Third, some are evolved stars such as blue horizontal branch stars or post-AGB stars \citep{tob87}. To better distinguish the nature of them, a medium or high spectral resolution observation is required.

\subsection{Summary}

In this work, we identify 22\,901 OB type spectra of 16\,032 stars from LAMOST DR5 dataset by filtering the EWs of spectral lines and manual inspection. A test with 5\,000 randomly selected subsamples shows that the identified OB samples can be complete to about $89\pm22$\% for OB stars earlier than B7 with $g$-band signal-to-noise ratio larger than $15$.

We apply MKCLASS to classify the identified OB stars into sub-classes and finally obtain sub-classes for 16\,782 OB stars. 89 are marked as ``Unclassifiable'' and 4\,922 are marked as question marks. These are likely because that the template spectra in MKCLASS may not match the observed spectra due to different metallicity or bad pixels in the observed spectra. An independent manual classification is conducted to 646 OB stars for validation. We find that the two results are quite similar with dispersion of about 1 sub-type in spectral sub-type and about 1 level in luminosity class. We then conclude that for most OB stars MKCLASS can give quite reliable results, while a few of them may be mis-classified.

The spatial distribution in Galactic coordinates shows that while most of the OB stars are located at low Galactic latitude. The spectral classes for the high Galactic latitude OB stars are substantially different with those located in the low Galactic latitude, which may either due to the observational selection effect or hint a different origin of the high Galactic latitude OB stars. These high Galactic latitude OB stars are particularly interesting in the sense that their origin is not clear. According to previous studies, some of these stars are likely the runaway stars scattered from the disk population and a few of them should be low-mass evolved BHB or post-AGB stars. It is worthy to follow up with high spectral resolution observations for these samples in future.

\acknowledgments

This work was supported by the National Key Basic Research Program of China 2014CB84570, the National Natural Science Foundation of China (NSFC) under grant 11773009, 11873057, 11390371, 11673007. The Guoshoujing Telescope (the Large Sky Area Multi-Object Fiber Spectroscopic Telescope LAMOST) is a National Major Scientific Project built by the Chinese Academy of Sciences. Funding for the project has been provided by the National Development and Reform Commission. LAMOST is operated and managed by the National Astronomical Observatories, Chinese Academy of Sciences. This research has made use of the SIMBAD database, operated at CDS, Strasbourg, France.

\software{MKCLASS (Gray \& Corbally 2014)}

\begin{table*} 
	\begin{center}
		\caption{OB stars identified in LAMOST DR5}\label{tab:OBstars}
		\resizebox{\textwidth}{!}{ 
		\begin{tabular}{c|c|c|c|c|c|c|c|c|c}
			\hline\hline
			Obsid & RA & Dec & S/N\tablenotemark{a}  & GroupID\tablenotemark{b}  & GroupSize\tablenotemark{c}  & SIMBAD & MKCLASS & Manual\tablenotemark{d} & Comments\tablenotemark{e}  \\
			\hline
250905055  &  0.059456  &  50.818608  &  82.53  &      &      &                        &    B9 II-III                     &    B                 &            \\
182612068  &  0.341024  &  50.132752  &  170.34  &      &      &                        &    kB9hF0mK0  Eu                 &    B                 &            \\
250509157  &  0.527828  &  39.183206  &  127.71  &      &      &    B9                  &    B9 V                          &    B                 &            \\
269504062  &  1.370397  &  57.288634  &  366.68  &      &      &    A2                  &    B9 V                          &    B                 &            \\
370703213  &  1.385054  &  46.492372  &  197.24  &      &      &                        &    B7 V                          &    B                 &            \\
364510227  &  2.630733  &  20.00764  &  256.31  &      &      &    M9                  &    B2 II                         &    B                 &            \\
178304113  &  2.63467  &  40.198498  &  824.28  &      &      &    B8                  &    A0 II                         &    B                 &            \\
266712010  &  5.5544246  &  3.5839873  &  32.48  &      &      &                        &                                  &                      &    sdOB    \\
171114077  &  5.5625166  &  56.5166045  &  113.43  &      &      &                        &    B9.5 II-III                   &    B                 &            \\
171102035  &  5.614378  &  54.270211  &  331.61  &      &      &                        &    kA0hF2mG5                     &    B                 &            \\
468710180  &  6.949154  &  34.674069  &  85.52  &  2945  &  2  &    sdOB                &                                  &                      &    sdOB    \\
75901223  &  6.9491834  &  34.674052  &  21.14  &  2945  &  2  &    sdOB                &                                  &                      &    sdOB    \\
171108164  &  8.4675215  &  55.1001548  &  135.23  &      &      &                        &    B5 II                         &    B3 V              &            \\
171109150  &  9.792189  &  56.185578  &  382.28  &      &      &                        &    B9 V                          &    B9.5  V           &            \\
469316204  &  9.922347  &  58.743764  &  468.34  &      &      &                        &    ?                             &    B                 &            \\
55601164  &  10.44285  &  25.82604  &  213.21  &  935  &  5  &    A3                  &    A9 mB9.5 V Lam Boo            &    B                 &            \\
59401164  &  10.44285  &  25.82604  &  160.12  &  935  &  5  &    A3                  &    A9 mB9 V Lam Boo              &    B                 &            \\
157501164  &  10.442872  &  25.826017  &  553.11  &  935  &  5  &    A3                  &    A9 mB7 V Lam Boo              &    B                 &            \\
157601164  &  10.442872  &  25.826017  &  609.77  &  935  &  5  &    A3                  &    A9 mB9 V Lam Boo              &    B                 &            \\
194411065  &  10.442872  &  25.826017  &  909.16  &  935  &  5  &    A3                  &    A9 mA0 V Lam Boo              &    B                 &            \\
353516249  &  15.840044  &  49.225157  &  27.18  &  2908  &  2  &                        &                                  &                      &    WD      \\
403216249  &  15.840044  &  49.225157  &  17.48  &  2908  &  2  &                        &                                  &                      &    WD      \\
368502216  &  18.444903  &  0.4746379  &  93.91  &      &      &    DO                  &                                  &                      &    WD      \\
392201217  &  19.8885832  &  39.6548095  &  222.77  &  1976  &  3  &                        &    B5 II                         &    B5 II             &            \\
392912106  &  19.8885832  &  39.6548095  &  200.05  &  1976  &  3  &                        &    B5 II                         &    B5 II             &            \\
190810208  &  19.88863  &  39.654762  &  242.13  &  1976  &  3  &                        &    B5 II                         &    B5 II             &            \\
380814123  &  26.899496  &  55.426933  &  346.78  &      &      &                        &    B1 V                          &    B2 V              &            \\
380810029  &  27.958356  &  53.846751  &  416.5  &      &      &                        &    B2 IV-V                       &    B3 III            &            \\
380812034  &  33.163063  &  56.548868  &  401.34  &      &      &                        &    B1 V                          &    B3 III            &            \\
180305005  &  35.715096  &  41.479483  &  23.95  &  1948  &  2  &                        &    kB8hA1mA4                     &    O                 &            \\
180405005  &  35.715712  &  41.480129  &  31.34  &  1948  &  2  &                        &    ?                             &    O                 &            \\
15307059  &  35.7392  &  57.009  &  135.03  &      &      &                        &    Unclassifiable                &    B                 &            \\
...& ...&... & ... &... &... & ...&...  & ...& ...\\
 			\hline\hline
		\end{tabular}}
	\end{center}
	\tablenotetext{a}{The signal to noise at $g$ band.}	
	\tablenotetext{b}{The ID number of the star for which was observed many times.}	
	\tablenotetext{c}{The number of exposures for a same star.}	
	\tablenotetext{d}{The spectral/luminosity types identified by eyes in this work.}
	\tablenotetext{e}{The ''peculiar" spectra identified by eyes.}
(This table is available in its entirety in a machine-readable form in the online journal. A portion is shown here for guidance regarding its form and content.)

\end{table*}

\end{document}